\documentclass[11pt,a4paper,twoside]{article} 
\usepackage{,amsmath,amssymb,latexsym,epsfig,floatflt,subfigure}
\newcommand{\scr}{\scriptscriptstyle}

\newcommand{\muk}{\mu^{\scr +} K^0}
\newcommand{\epi}{e^{\scr +} \pi^0}
\newcommand{\et}{e^{\scr +} \eta}
\newcommand{\ek}{e^{\scr +} K^0}
\newcommand{\mpi}{\mu^{\scr +} \pi^0}
\newcommand{\mt}{\mu^{\scr +} \eta}
\newcommand{\ero}{e^{\scr +} \rho^0}
\newcommand{\eo}{e^{\scr +} \omega}
\newcommand{\eks}{e^{\scr +} K^{*0}}
\newcommand{\mro}{\mu^{\scr +} \rho^0}
\newcommand{\mo}{\mu^{\scr +} \omega}
\newcommand{\nep}{\nu^{\scr C}_{\scr e} \pi^+}
\newcommand{\nek}{\nu^{\scr C}_{\scr e} K^+}
\newcommand{\nmpi}{\nu^{\scr C}_{\scr \mu} \pi^+}
\newcommand{\nmk}{\nu^{\scr C}_{\scr \mu} K^+}
\newcommand{\nero}{\nu^{\scr C}_{\scr e} \rho^+}
\newcommand{\neks}{\nu^{\scr C}_{\scr e} K^{*+}}
\newcommand{\nmro}{\nu^{\scr C}_{\scr \mu} \rho^+}
\newcommand{\nmks}{\nu^{\scr C}_{\scr \mu} K^{*+}}
\newcommand{\ntp}{\nu^{\scr C}_{\scr \tau} \pi^+}
\newcommand{\ntk}{\nu^{\scr C}_{\scr \tau} K^+}
\newcommand{\ntro}{\nu^{\scr C}_{\scr \tau} \rho^+}
\newcommand{\ntks}{\nu^{\scr C}_{\scr \tau} K^{*+}}
\newcommand{\epin}{e^{\scr +} \pi^-}
\newcommand{\mpin}{\mu^{\scr +} \pi^-}
\newcommand{\eron}{e^{\scr +} \rho^-}
\newcommand{\mron}{\mu^{\scr +} \rho^-}
\newcommand{\nepn}{\nu^{\scr C}_{\scr e} \pi^0}
\newcommand{\nekn}{\nu^{\scr C}_{\scr e} K^0}
\newcommand{\nmpin}{\nu^{\scr C}_{\scr \mu} \pi^0}
\newcommand{\nmkn}{\nu^{\scr C}_{\scr \mu} K^0}
\newcommand{\neron}{\nu^{\scr C}_{\scr e} \rho^0}
\newcommand{\neksn}{\nu^{\scr C}_{\scr e} K^{*0}}
\newcommand{\nmron}{\nu^{\scr C}_{\scr \mu} \rho^0}
\newcommand{\nmksn}{\nu^{\scr C}_{\scr \mu} K^{*0}}
\newcommand{\neon}{\nu^{\scr C}_{\scr e} \omega}
\newcommand{\nmon}{\nu^{\scr C}_{\scr \mu} \omega}
\newcommand{\netn}{\nu^{\scr C}_{\scr e} \eta}
\newcommand{\nmtn}{\nu^{\scr C}_{\scr \mu} \eta}
\newcommand{\ntpn}{\nu^{\scr C}_{\scr \tau} \pi^0}
\newcommand{\ntkn}{\nu^{\scr C}_{\scr \tau} K^0}
\newcommand{\nttn}{\nu^{\scr C}_{\scr \tau} \eta}
\newcommand{\ntron}{\nu^{\scr C}_{\scr \tau} \rho^0}
\newcommand{\nton}{\nu^{\scr C}_{\scr \tau} \omega}
\newcommand{\ntksn}{\nu^{\scr C}_{\scr \tau} K^{*0}}
\begin{document}
\def\NPB#1#2#3{Nucl.\ Phys.\ {\bf B}\,{\bf #1} (19#2) #3}
\def\PLB#1#2#3{Phys.\ Lett.\ {\bf B}\,{\bf #1} (19#2) #3}
\def\PRD#1#2#3{Phys.\ Rev.\ {\bf D}\,{\bf #1} (19#2) #3}
\def\PRL#1#2#3{Phys.\ Rev.\ Lett.\ {\bf#1} (19#2) #3}
\def\PRT#1#2#3{Phys.\ Rep.\ {\bf#1}\,{\bf C} (19#2) #3}
\def\ARAA#1#2#3{Ann.\ Rev.\ Astron.\ Astrophys.\ {\bf#1} (19#2) #3}
\def\ARNP#1#2#3{Ann.\ Rev.\ Nucl.\ Part.\ Sci.\ {\bf#1} (19#2) #3}
\def\MODA#1#2#3{Mod.\ Phys.\ Lett.\ {\bf A}\,{\bf#1} (19#2) #3}
\def\NC#1#2#3{Nuovo\ Cim.\ {\bf#1} (19#2) #3}
\def\ANPH#1#2#3{Ann.\ Phys.\ {\bf#1} (19#2) #3}
\def\PROG#1#2#3{Prog.\ Theor.\ Phys.\ {\bf#1} (19#2) #3}
\parindent 0pt

\begin{titlepage}
\thispagestyle{empty}
\parindent 0pt
\title{{\bf Large Quark Rotations, Neutrino mixing \\ and Proton Decay}}
\author{
Yoav Achiman\footnote{e-mail: achiman@theorie.physik.uni-wuppertal.de}
$\dagger$
\ and
Carsten Merten\footnote{e-mail: merten@theorie.physik.uni-wuppertal.de}\\
[0.5cm]
        Department of Physics\\
        University of Wuppertal\\
        Gau\ss{}str.~20, D--42097 Wuppertal \\
        Germany\\
        $\dagger$ School of Physics and Astronomy\\
        Tel Aviv University\\
        69978  Tel Aviv, Israel\\ [1.5cm]}

\date{March 2000}

\maketitle
\setlength{\unitlength}{1cm}
\begin{picture}(5,1)(-12.5,-12)
\put(0,0){WUB 00-03}
\end{picture}
\begin{picture}(5,1)(-7.4,-11.5)
\put(0,0){TAUP 2627-2000}
\end{picture}
\parindent0cm
\begin{abstract}
\noindent Right-handed (RH) rotations do not play a role in the 
Standard Model, and only the differences of the LH mixing angles are 
involved in ${\bf V}_{\scr \textrm{CKM}}$. This leads to the huge 
freedom in the fermionic mass matrices. However, that is no more true in 
extensions of the Standard Model. For example in GUTs large RH rotations of 
the quarks can be related to the observed large neutrino mixing or in 
particular, all mixing angles are relevant for the proton decay. We 
present a simple realistic non-SUSY $SO(10)$ GUT with large RH and LH mixing 
and study the corresponding nucleon decay rates.
\end{abstract}
\thispagestyle{empty}
\setcounter{page}{0}

\end{titlepage}

\section{Introduction}

What is the origin of the fermionic masses?
This is one of the open questions in the Standard Model (SM). 
The mass matrix entries are arbitrary parameters of
the model and only the neutrinos are restricted to be massless.

One can add conjectures for the structure of the fermionic mass
matrices to the SM. Many different conjectures are known 
to give the right masses of the charged fermions
and ${\bf V}_{\scr \textrm{CKM}}$ (within the experimental errors) 
and this is clearly an indication that the mass problem is far 
from being solved.

The main reason for this large freedom is that right-handed (RH)
rotations~\footnote{A general non-hermitian matrix is diagonalized by
  a bi-unitary transformation. This means that two unitary matrices,
  one from the left and one from the right, are needed. Those matrices
  are equal only for hermitian (or symmetric) matrices~\cite{corf}.}
are non-observable in the SM. Also the observed left-handed
(LH) mixing matrix ${\bf V}_{\scr \textrm{CKM}}$ involves only the 
{\em differences} between the mixing angles of the $u$-like and
$d$-like quarks and the individual mixing can be large. Actually,
there is already a strong indication from the neutrino sector that 
large rotations are required~\cite{nu1,nu2}.

The considerable freedom in the mass matrices of the SM does not exist
in its extensions. We know that the SM must be extended for many
reasons if not alone to explain the fermionic masses and mixing angles. In
particular, in its  most popular extension, the Grand Unified Theories
(GUTs)~\cite{gut}, all mixing angles are relevant, as will be explained later.
In such a framework one cannot change the ``weak basis'' without
changing the physics.

The aim of this paper is to present a simple realistic $SO(10)$ GUT~\cite{s10}
model with large LH as well as RH mixing angles and to use them to
study the consequences for nucleon decay. Mixing effects were
generally neglected in the conventional proton decay models~\cite{pd} 
by assuming that the mixing angles are small.

Large rotations in the quark sector could be a natural reason for the 
large mixing observed in the leptonic sector in terms of neutrino
oscillations. In particular, there is a kind of duality between the RH 
mixing of the quarks and the leptonic LH rotations~\cite{du}.

A predictive model for the fermionic masses must involve a family
symmetry which dictates the texture of the mass matrices and protects
it from getting large radiative corrections. We used a global 
$U(1)_{\scr F}$~\cite{u1} in the framework of an $SO(10)$ GUT that 
will add relations between the matrix elements. Our aim was to 
look for the simplest possible realization of a realistic model 
with large mixing angles. We therefore used the famous $SO(10)$ paper 
of Harvey, Reiss and Ramond~\cite{hrr} and generalized it to
asymmetric matrices. Such matrices give usually large LH and/or RH
mixing angles.

We did not use non-renormalizable contributions to the mass matrices
\`{a} la Froggatt and Nielsen~\cite{fn} because this method assumes
ad hoc physics beyond the GUT and many new particles. In addition, 
the resulting matrix elements are given there in orders of magnitude
only. This can explain the hierarchy of the masses but not the light 
see-saw neutrino properties. Those are obtained from a product of
three matrices and hence predicted up to a factor of $[\mathcal{O}(1)]^3$ 
which may be quite large~\cite{capri}.

Most recent models use SUSY GUTs~\cite{sgut}~\footnote{We shall use it 
also in a forthcoming paper~\cite{ar}.}. However, the available 
parameter space of low-energy SUSY shrinked recently so much that MSSM 
is on the verge of loosing its ``naturality''~\cite{nat}. At the same time 
solutions without low energy SUSY have emerged quite naturally~\cite{sM} in
superstring and M theory, and also the hierarchy problem can be solved 
adding extra dimensions~\cite{dim}. We think therefore that it is
worthwhile to look for a non-SUSY GUT which is consistent with all
observed experimental facts. The hope is that the fine tuning
(hierarchy) problem will be solved in the more fundamental theory. 
Note also that conventional GUTs are relatively simple and give much 
more reliable predictions for nucleon decay than SUSY theories. 

Unification of the gauge coupling constants is obtained using an
intermediate breaking scale~\cite{int} $M_{\scr I} \simeq 10^{11}$ GeV. 
This is very useful because $M_{\scr I}$ is also the right mass scale
for the RH neutrinos needed for the see-saw mechanism as well as for
leptogenesis as the origin of the baryon asymmetry~\cite{lba} and 
the (invisible) axion window~\cite{ax}.

The observed neutrino oscillations teach us about the neutrino masses
and mixing angles. This is the first evidence for physics beyond the
SM. Our claim is that observation of other phenomena like RH currents, 
leptoquarks, baryon asymmetry induced by leptogenesis and especially 
nucleon decay can reveal the unknown mixing angles and reduce considerably
the freedom in the fermionic mass matrices.

The plan of the paper is as follows. In section 2 we discuss the
symmetry breaking that is dictated by the requirement of gauge
unification and the Higgs representations needed to give the correct fermion
masses. The mass matrices and our global $U(1)_{\scr F}$ details are given
in section 3. In section 4 the numerical solutions for the mass
matrices are obtained by the use of the renormalization group
equations (RGEs), and a fit to the observed properties of the charged
fermions is elaborated. Three solutions are found which are consistent
with the observed neutrino anomalies~\cite{nu1,nu2,nu,chooz} (except 
for LSND~\cite{LSND}). All mixing angles for those solutions are 
obtained explicitly and this allows for the calculation of the nucleon 
decay rates in section 5. Section 6 is devoted to the conclusions.   
 
\section{Symmetry breaking of the $SO(10)$ GUT}
 
We use an $SO(10)$ GUT~\cite{s10} which is broken down to the SM
via an intermediate scale $M_{\scr I}$ as follows 
\begin{equation}
SO(10) \; \stackrel{M_{\scr U}}{\longrightarrow} \; 
G_{\scr I} \; \stackrel{M_{\scr I}}{\longrightarrow} \; 
G_{\scr \textrm{SM}} \; \stackrel{M_{\scr Z}}{\longrightarrow} \; 
SU(3)_{\scr C} \otimes U(1)_{\scr \textrm{em}} \; ,
\end{equation}
where the intermediate symmetry group is the Pati-Salam one~\cite{ps},
$G_{\scr I} =  G_{\scr \textrm{PS}} \equiv SU(4)_{\scr C} 
\otimes SU(2)_{\scr L} \otimes SU(2)_{\scr R}$.

The breaking at $M_{\scr U}$ is done using the Higgs representations 
$\Phi_{\bf{210}}$ (or $\Phi_{\bf{54}}$ in models with $D$ parity) 
while for the breaking at $M_{\scr I}$ we use the SM singlet of a 
$\Phi_{\bf{126}}$, i.e. the $G_{\scr \textrm{PS}}$ representation 
$(\bf{10,1,3})_{\bf{126}}$. For the masses of the Higgs scalars the 
``extended survival hypothesis''~\cite{esh} is assumed.

In view of the representation content of the mass terms
\begin{equation}
{\bf 16} \otimes {\bf 16} \; = \; ({\bf 10} \oplus {\bf 126})_{\textrm{symm}}
\oplus {\bf 120}_{\textrm{antisymm}}
\end{equation}
we give the light fermions masses via the VEVs of
$(\bf{1,2,2})_{\bf 10/120}$ and $(\bf{15,2,2})_{\bf 120/126}$.  
The RH neutrino masses of order $M_{\scr R} \sim M_{\scr I}$ will
be induced via the VEV of the above mentioned $(\bf{10,1,3})_{\bf{126}}$.

The Higgs doublet of the SM is therefore a linear combination of the
$SU(2)_{\scr L}$ doublets in the representations $(\bf{1,2,2})$ and 
$(\bf{15,2,2})$. The exact number of representations needed for the 
fermion mass matrices will be given later when we will discuss those 
matrices. This number is needed however to fix the RGEs used to
calculate $M_{\scr I}$, $M_{\scr U}$ and
$\alpha_{\scr U}(M_{\scr U})$ as well as the values of the 
mass matrices at $M_{\scr I}$~\footnote{Note that due to the
quark-lepton symmetry of $G_{\scr \textrm{PS}}$ the $SO(10)$ mass
relations are valid also at the scale $M_{\scr I}$.}.
As will be explained in the next chapter, we used in the
RGEs $N_1=$ number of $({\bf 1,2,2})=4$ and $N_{15}=$ number of 
$({\bf 15,2,2})=2$ as well as 
$\Delta_L=$ number of $({\bf \overline{10},3,1})_{\bf 126}=0$ and
$\Delta_R=$ number of $({\bf 10,1,3})_{\bf 126}=1$ for the 
$G_{\scr \textrm{PS}}$ symmetry breaking.

For the matching conditions at $M_{\scr I}$ we took
\begin{eqnarray} \label{psmc1}
    \alpha^{-1}_{\scr 4C}(M_{\scr I}) & = & \alpha^{-1}_{\scr 3}(M_{\scr I})
    + \frac{1}{12\pi} \\ \label{psmc2}
    \alpha^{-1}_{\scr 2L}(M_{\scr I}) & = & \alpha^{-1}_{\scr 2}(M_{\scr I}) 
    \\ \label{psmc3}
    \alpha^{-1}_{\scr 2R}(M_{\scr I}) & = & 
    \frac{5}{3} \, \alpha^{-1}_{\scr 1}(M_{\scr I}) - 
    \frac{2}{3} \, \alpha^{-1}_{\scr 3}(M_{\scr I}) + \frac{1}{3\pi}
  \end{eqnarray}
while  at $M_{\scr U}$ the conditions are\\
  \begin{eqnarray}
    \alpha^{-1}_{\scr U}(M_{\scr U}) & = & \alpha^{-1}_{\scr 4C}(M_{\scr U})
    + \frac{1}{3\pi} \\
    & = & \alpha^{-1}_{\scr 2L}(M_{\scr U}) + \frac{1}{2\pi} \\
    & = & \alpha^{-1}_{\scr 2R}(M_{\scr U}) + \frac{1}{2\pi}
  \end{eqnarray}
In general the matching conditions between gauge couplings belonging to
theories with symmetry groups ${\bf \mathcal{G}}_{j}$ and 
${\bf \mathcal{G}}_{k}$ can be written as
\begin{equation}
    \alpha^{-1}_{\scr j}(M^{}_{\scr I/U}) - \dfrac{1}{12\pi} \,
    S_2({\bf \mathcal{G}}_{j}) \; = \; \alpha^{-1}_{\scr k}(M^{}_{\scr I/U}) 
    - \dfrac{1}{12\pi} \, S_2({\bf \mathcal{G}}_{k})
  \end{equation}
where $S_2({\bf \mathcal{G}}_{j})$ is the Dynkin index of the adjoint
representation of the group $G_j$. Details can be found in~\cite{cm},
the results are given in Table \ref{int}.
\begin{table}[h]
\begin{center}
\begin{tabular}{|l||c|c|c|}
\hline Quantity & $M_{\scr I}$ & $\alpha_{\scr 1}(M_{\scr I})$ 
& $ \alpha_{\scr 2}(M_{\scr I})$ \\
\hline Value & $6.14 \cdot 10^{10}$ GeV & $(46.00)^{-1}$ 
& $(39.86)^{-1}$  \\
\hline \hline
Quantity & $\alpha_{\scr 3}(M_{\scr I})$ & $\alpha_{\scr 2R}(M_{\scr I})$ 
& $\alpha_{\scr 2L}(M_{\scr I})$ \\
\hline Value & $(31.39)^{-1}$ & $(55.85)^{-1}$ &  $(39.86)^{-1}$ \\
\hline \hline
Quantity & $\alpha_{\scr 4C}(M_{\scr I})$ & $M_{\scr U}$ 
& $\alpha_{\scr U}(M_{\scr I})$ \\
\hline Value &$(31.42)^{-1}$  &  $1.31 \cdot 10^{16}$ GeV & $(20.08)^{-1}$ \\
\hline
\end{tabular}
\end{center}
\caption{ Symmetry breaking scales and gauge couplings
for $N_1=4$ and $N_{15}=2$ \label{int}}
\end{table}
We checked also the RGEs for the intermediate symmetry  
$G_{\scr I}= G_{\scr \textrm{PS}} \otimes D$, where $D$ is the 
discrete $D$-parity which requires $\alpha_{\scr 2L}=\alpha_{\scr 2R}$
between $M_{\scr U}$ and $M_{\scr I}$. In this case however $M_{\scr U}
(G_{\scr {\scr \textrm{PS}}}\otimes D)=1.16 \cdot
10^{15}$  GeV, a value which would lead to a too fast proton decay.

\section{The mass matrices}
 
The aim of this paper is to present models with large RH and LH
mixing angles of the quarks that lead to large mixing of the 
leptons and to study the predictions of those models for 
the nucleons decay ratios. To generate the mass matrices we use the 
method of Harvey, Reiss and Ramond~\cite{hrr} as it was realized in~\cite{ag} 
for symmetric mass matrices. However, in order to have large mixing angles
one should rather use asymmetric mass matrices. For asymmetric 
matrices we need to make certain changes in the above scenario
and add also the antisymmetric Higgs representation $\Phi_{\bf{120}}$.
Practically speaking , we shall use a global family symmetry
$U(1)_{\scr F}$ or $Z_n$ that will dictate the neutrino properties in 
terms of the observed masses and mixings of the charged fermions. This 
symmetry will be chosen in such a way that the predictive Fritzsch 
texture~\cite{fri} will be realized. However, as it is well known the 
symmetric version of this texture cannot account for the large top 
quark mass~\cite{top}. As we need anyhow asymmetric mass matrices  
we shall use the asymmetric Fritzsch texture which is also known under 
the name ``nearest neighbour interaction'' model (NNI)~\cite{nni}. Namely,
\begin{equation} \label{mmans}
{\bf M} \; = \; \left( \begin{array}{lll}
0 & A & 0 \\ B & 0 & C \\ 0 & D & E
\end{array} \right) \, .
\end{equation}
In view of the fact that we are actually mainly interested in the 
predictions for the nucleon decay rates which are not sensitive to 
the details of $CP$ violation we will use for simplicity real mass 
matrices~\footnote{This can help to solve the strong $CP$ problem while 
the observed $CP$ violation in the $K$-decay can come from a different 
origin than the CKM matrix.}.  

The three fermion families and the different Higgs representations in  
${\bf 16}_i {\bf {\overline \Phi}}_k {\bf 16}_j$ transform under the global 
$U(1)_{\scr F}$ as follows
\begin{eqnarray}
{\bf 16}_j & \rightarrow & \exp(i \alpha_j \theta){\bf 16}_j \\
{\bf \Phi}_k & \rightarrow & \exp(i \beta_k \theta) {\bf \Phi}_k
\end{eqnarray}
The invariance under $U(1)_{\scr F}$ requires therefore that the 
$\beta_k$ must obey $\alpha_i + \alpha_j = \beta_k$ . Hence, the 
fermionic part of the mass matrices has the following quantum numbers:
\begin{equation}
{\bf M}_f \; \sim \; \left( \begin{array}{lll}
\alpha_1+\alpha_1 & \alpha_1+\alpha_2 & \alpha_1+\alpha_3 \\
\alpha_1+\alpha_2 & \alpha_2+\alpha_2 & \alpha_2+\alpha_3 \\ 
\alpha_1+\alpha_3 & \alpha_2+\alpha_3 & \alpha_3+\alpha_3
\end{array} \right).
\end{equation}
To realize the NNI texture (\ref{mmans}) one sees that only 
Higgs representations with the charges 
$\beta = \alpha_1 + \alpha_2 , \alpha_2 + \alpha_3$ and 
$ \alpha_3 + \alpha_3$ can couple to the fermions. Also, we still have the
possibility to couple one Higgs representation to two different
combinations i.e. $\alpha_1 + \alpha_2 = 2\alpha_3$.

Taking all this into account the Yukawa coupling matrices (at energies
$\mu \gtrsim M_{\scr I}$) can have the structure
\begin{eqnarray} \nonumber
{\bf Y}_{\bf \scr 10}^{\scr (1)} = \begin{pmatrix}
0 & x_1 & 0 \\ x_1 & 0 & 0 \\ 0 & 0 & \tilde x_1 \end{pmatrix} \; ; \quad
{\bf Y}_{\bf \scr 126}^{\scr (1)} = \begin{pmatrix}
0 & y_1 & 0 \\ y_1 & 0 & 0 \\ 0 & 0 & \tilde y_1 \end{pmatrix} \; ; \quad
{\bf Y}_{\bf \scr 120}^{\scr (1)} = \begin{pmatrix}
0 & z_1 & 0 \\ -z_1 & 0 & 0 \\ 0 & 0 & 0 \end{pmatrix} \; ; & & \\ \label{ycm}
{\bf Y}_{\bf \scr 10}^{\scr (2)} = \begin{pmatrix}
0 & 0 & 0 \\ 0 & 0 & x_2 \\ 0 & x_2 & 0 \end{pmatrix} \hspace{0.28cm} ; \quad
{\bf Y}_{\bf \scr 126}^{\scr (2)} = \begin{pmatrix}
0 & 0 & 0 \\ 0 & 0 & y_2 \\ 0 & y_2 & 0 \end{pmatrix} \hspace{0.22cm} ; \quad
{\bf Y}_{\bf \scr 120}^{\scr (2)} = \begin{pmatrix}
0 & 0 & 0 \\ 0 & 0 & z_2 \\ 0 & -z_2 & 0 \end{pmatrix} \hspace{0.25cm} & &
\end{eqnarray}
On top of that, we need at least one $\Phi_{\bf{126}}$ with a large
VEV in $ ({\bf  10, 1, 3})$ to break the $SO(10)$ gauge symmetry at
$M_{\scr I}$. This VEV will generate also the RH neutrinos masses 
$M_{\scr R} \sim M_{\scr I}$. Actually, as is discussed in App.\,I  
we will use only one $\Phi_{\bf{126}}$ in addition to two $\Phi_{\bf{10}}$ 
and two $\Phi_{\bf{120}}$ to generate the asymmetrical mass matrices. 
The detailed fits required four VEVs in the direction $({\bf 1,2,2})$  
and  two in that of $ ({\bf 15,2,2})$. Those give also the right 
unification as discussed before.
 
In terms of the notation and discussion of App.\,I we obtained the 
following expressions for the mass matrix elements in terms of the
VEVs and Yukawa couplings:
\begin{eqnarray} \label{dmm12}
({\bf M}_d)_{\scr 12} & = & 
x_1 \upsilon^{\scr (1)}_d + y_1 \omega^{\scr (1)}_d    
+ z_1 \tilde \upsilon^{\scr (1)}_d  
\\ \label{dmm21}
({\bf M}_d)_{\scr 21} & = & 
x_1 \upsilon^{\scr (1)}_d + y_1 \omega^{\scr (1)}_d 
- z_1 \tilde \upsilon^{\scr (1)}_d  
\\ \label{dmm23}
({\bf M}_d)_{\scr 23} & = & x_2 \upsilon^{\scr (2)}_d 
+ z_2 (\tilde \upsilon^{\scr (2)}_d + \tilde \omega^{\scr (2)}_d) 
\\ \label{dmm32}
({\bf M}_d)_{\scr 32} & = & x_2 \upsilon^{\scr (2)}_d 
- z_2 (\tilde \upsilon^{\scr (2)}_d + \tilde \omega^{\scr (2)}_d) \\ \nonumber
({\bf M}_d)_{\scr 33} & = & \tilde x_1 \upsilon^{\scr (1)}_d 
+ \tilde y_1 \omega^{\scr (1)}_d \\ \label{dmm33}
       & = & \Big( \dfrac{\tilde x_1}{x_1} \Big) x_1 \upsilon^{\scr (1)}_d
           + \Big( \dfrac{\tilde y_1}{y_1} \Big) y_1 \omega^{\scr (1)}_d
\\ \nonumber & & \\ \nonumber
({\bf M}_e)_{\scr 12} & = & 
x_1 \upsilon^{\scr (1)}_d - 3 \, y_1 \omega^{\scr (1)}_d 
+ z_1 \tilde \upsilon^{\scr (1)}_d  
\\ \label{emm12}
& = & ({\bf M}_d)_{\scr 12} -4 \, y_1 \omega^{\scr (1)}_d  \\ \nonumber
({\bf M}_e)_{\scr 21} & = & 
x_1 \upsilon^{\scr (1)}_d - 3 \, y_1 \omega^{\scr (1)}_d 
- z_1 \tilde \upsilon^{\scr (1)}_d  
\\ \label{emm21}
& = & ({\bf M}_d)_{\scr 21} -4 \, y_1 \omega^{\scr (1)}_d \\ \nonumber
({\bf M}_e)_{\scr 23} & = & 
x_2 \upsilon^{\scr (2)}_d + z_2 (\tilde \upsilon^{\scr (2)}_d 
-3 \, \tilde \omega^{\scr (2)}_d) \\ \label{emm23}
& = & ({\bf M}_d)_{\scr 23} -4 \, z_2\tilde \omega^{\scr (2)}_d 
\\ \nonumber
({\bf M}_e)_{\scr 32} & = & 
x_2 \upsilon^{\scr (2)}_d - z_2 (\tilde \upsilon^{\scr (2)}_d 
-3 \, \tilde \omega^{\scr (2)}_d) \\ \label{emm32}
& = & ({\bf M}_d)_{\scr 32} +4 \, z_2\tilde \omega^{\scr (2)}_d \\ \nonumber
({\bf M}_e)_{\scr 33} & = & \tilde x_1 \upsilon^{\scr (1)}_d 
- 3 \, \tilde y_1 \omega^{\scr (1)}_d \\ \label{emm33}
& = & ({\bf M}_d)_{\scr 33} 
-4 \, \Big( \dfrac{\tilde y_1}{y_1} \Big) y_1  \omega^{\scr (1)}_d
\\ \nonumber & & \\ \label{umm12}
({\bf M}_u)_{12} & = & x_1 \upsilon^{\scr (1)}_u + y_1 \omega^{\scr (1)}_u 
+ z_1 \tilde \upsilon^{\scr (1)}_u  
\\ \label{umm21}
({\bf M}_u)_{21} & = & x_1 \upsilon^{\scr (1)}_u + y_1 \omega^{\scr (1)}_u 
- z_1 \tilde \upsilon^{\scr (1)}_u  
\\ \label{umm23}
({\bf M}_u)_{23} & = & x_2 \upsilon^{\scr (2)}_u 
+ z_2 (\tilde \upsilon^{\scr (2)}_u + \tilde \omega^{\scr (2)}_u) 
\\ \label{umm32}
({\bf M}_u)_{32} & = & x_2 \upsilon^{\scr (2)}_u 
- z_2 (\tilde \upsilon^{\scr (2)}_u + \tilde \omega^{\scr (2)}_u) \\ \nonumber
({\bf M}_u)_{33} & = & \tilde x_1 \upsilon^{\scr (1)}_u 
+ \tilde y_1 \omega^{\scr (1)}_u \\ \label{umm33}
       & = & \Big( \dfrac{\tilde x_1}{x_1} \Big) x_1 \upsilon^{\scr (1)}_u
           + \Big( \dfrac{\tilde y_1}{y_1} \Big) y_1 \omega^{\scr (1)}_u
\\ & & \nonumber \\ \nonumber
({\bf M}^{\scr (\textrm{Dir})}_\nu)_{12} & = & 
x_1 \upsilon^{\scr (1)}_u - 3 \, y_1 \omega^{\scr (1)}_u 
+ z_1 \tilde \upsilon^{\scr (1)}_u  
\\ \label{nmm12}
& = & ({\bf M}_u)_{12} -4 \, y_1 \omega^{\scr (1)}_u  \\ \nonumber
({\bf M}^{\scr (\textrm{Dir})}_\nu)_{21} & = & 
x_1 \upsilon^{\scr (1)}_u - 3 \, y_1 \omega^{\scr (1)}_u 
- z_1 \tilde \upsilon^{\scr (1)}_u  
\\ \label{nmm21}
& = & ({\bf M}_u)_{21} -4 \, y_1 \omega^{\scr (1)}_u  \\ \nonumber 
({\bf M}^{\scr (\textrm{Dir})}_\nu)_{23} & = & 
x_2 \upsilon^{\scr (2)}_u + z_2 (\tilde \upsilon^{\scr (2)}_u 
-3 \, \tilde \omega^{\scr (2)}_u) \\ \label{nmm23}
& = & ({\bf M}_u)_{23} -4 \, z_2\tilde \omega^{\scr (2)}_u \\ \nonumber
({\bf M}^{\scr (\textrm{Dir})}_\nu)_{32} & = & 
x_2 \upsilon^{\scr (2)}_u - z_2 (\tilde \upsilon^{\scr (2)}_u 
-3 \, \tilde \omega^{\scr (2)}_u) \\ \label{nmm32}
& = & ({\bf M}_u)_{32} +4 \, z_2\tilde \omega^{\scr (2)}_u 
\end{eqnarray}
\begin{eqnarray} \nonumber
({\bf M}^{\scr (\textrm{Dir})}_\nu)_{33} & = & 
\tilde x_1 \upsilon^{\scr (1)}_u - 3 \, \tilde y_1 \omega^{\scr (1)}_u 
\\ \label{nmm33}
& = & ({\bf M}_u)_{33} -4 \, \Big( \dfrac{\tilde y_1}{y_1} \Big) 
y_1 \omega^{\scr (1)}_u
\end{eqnarray}
All those elements are obtained from 14 independent Higgs parameters
which are products of VEVs and Yukawa couplings. They also fix the 
matrix elements of the RH neutrino Majorana mass matrix (in terms of
the VEV of $({\bf 10,1,3})$ that breaks 
$G_{\scr \textrm{PS}} \rightarrow G_{\scr \textrm{SM}}$), with 
$M_{\scr R} \sim M_{\scr I}$ being a quasi-free parameter:
\begin{equation} \label{rhmm}
{\bf M}^{\scr (\textrm{Maj})}_{\nu \scr{R}} \; = \; M_{\scr R} \,
\begin{pmatrix}
0 & y_1 & 0 \\ y_1 & 0 & 0 \\ 0 & 0 & \tilde y_1 \end{pmatrix}
\; = \; y_1 \, M_{\scr R} \, \begin{pmatrix}
0 & 1 & 0 \\ 1 & 0 & 0 \\ 0 & 0 & \tilde y_1/y_1 \end{pmatrix}
\end{equation}
By diagonalizing the charged fermion mass matrices 
\begin{eqnarray} \nonumber
{\bf U}^{\dagger}_{\scr L} \, {\bf M}^{}_u \, {\bf U}^{}_{\scr R} & = &
{\bf M}^{\scr (D)}_u \; , \quad
{\bf D}^{\dagger}_{\scr L} \, {\bf M}^{}_d \, {\bf D}^{}_{\scr R} \ = \
{\bf M}^{\scr (D)}_d \; , \\ \label{soeq}
{\bf E}^{\dagger}_{\scr L} \, {\bf M}^{}_e \, {\bf E}^{}_{\scr R} \ & = &
{\bf M}^{\scr (D)}_e \; , \quad
{\bf U}^\dagger_{\scr L} {\bf D}^{}_{\scr L} \ = \ 
{\bf V}_{\scr \textrm{CKM}} \; ,
\end{eqnarray}
and the light see-saw neutrino mass matrix
\begin{equation}
{\bf M}_\nu^{\scr \textrm{light}} \; \simeq \; -
{\bf M}^{\scr (\textrm{Dir})}_\nu \,
\big( {\bf M}^{\scr (\textrm{Maj})}_{\nu \scr{R}} \big)^{-1}
\big( {\bf M}^{\scr (\textrm{Dir})}_\nu \big)^T
\end{equation}
we  obtained all masses of the charged fermions and those of the 
three light neutrinos as well as the mixing matrices ${\bf U}_{\scr L,R}$, 
${\bf D}_{\scr L,R}$, ${\bf E}_{\scr L,R}$ and 
${\bf N}_\nu$~\footnote{There is only one neutrino mixing matrix 
as ${\bf M}_\nu^{\scr \textrm{light}}$ is always symmetric.}.

Neglecting phases (as explained before) we have 12 masses and 21 
mixing parameters. All those fix the proton decay rates and other GUT 
scale effects (like baryon asymmetry induced by leptogenesis~\cite{lba}). 
However, in the framework of the SM (with massive neutrinos) the only
observable mixing matrices would be
${\bf V}_{\scr \textrm{CKM}}= {\bf U}_{\scr L}^{\dagger} {\bf D}_{\scr L}$ 
and  ${\bf U} = {\bf E}_{\scr L}^{\dagger} {\bf N}_\nu$. This 
gives 18 mixing observables~\footnote{Note however, given the 
mass matrices we have definite predictions for the proton decay, 
the baryon asymmetry and RH currents which can be measured in principle 
by future experiments.}.

\section{Numerical solutions for the mass matrices}

The first step will be to calculate numerically the mass and mixing
matrices in (\ref{soeq}). This is done using the $SO(10)$ relations
(\ref{dmm12}-\ref{nmm33}) for the mass matrices at $\mu = M_{\scr I} 
= 6.14 \cdot 10^{10}$ GeV. On the RHS of (\ref{soeq}) we use the 
calculated masses of the charged fermions (see Table~\ref{FermMass2}) 
and the (real) CKM matrix at $\mu = M_{\scr I}$. Those values are 
obtained using the RGEs as is described in detail in~\cite{cm}. We
then run the mass matrices from $M_{\scr I}$ to $M_{\scr Z}$,
diagonalize them there and compare the obtained masses and mixings 
with their experimentally observed values.
\begin{table}[h]
\begin{center}
\begin{tabular}{|l||c|c|c|}
\hline Mass & $m_u(M_{\scr I})$ & $m_d(M_{\scr I})$ & $m_s(M_{\scr I})$ \\
\hline Value & $1.16$ MeV & $2.38$ MeV & $47.4$ MeV \\
\hline \hline
Mass & $m_c(M_{\scr I})$ & $m_b(M_{\scr I})$ & $m_t(M_{\scr I})$ \\
\hline Value & $337.6$ MeV & $1360$ MeV & $101.2$ GeV \\
\hline \hline
Mass & $m_e(M_{\scr I})$ & $m_\mu(M_{\scr I})$ & $m_\tau(M_{\scr I})$ \\
\hline Value & $513$ keV & $108.14$ MeV & $1838.3$ MeV \\
\hline
\end{tabular}
\end{center}
\caption{\label{FermMass2} Fermion masses at $M_{\scr I}$}
\end{table}

Now, without using the neutrino sector, only 13 of our 14 parameters are
independent because only the combination 
$z_2 (\tilde \upsilon^{\scr(2)}_u + \tilde \omega^{\scr (2)}_u)$    
appears in the charged fermion equations. We have in (\ref{soeq}) 30  
nonlinear equations. On the other hand each of the (real) 6 mixing 
matrices is parametrized by 3 angles, so altogether there are 18 mixing
angles. With the 13 Higgs parameters we have 31 ``unknowns''. One of
them must therefore be ``given'' to be able to search for solutions 
numerically. For the given quantity we use the ratio $\tilde y_1/y_1$ 
which determines the RH neutrino mass matrix (\ref{rhmm}) except 
for a global factor and we vary its 
value between $1 \leq |\tilde y_1/y_1| \leq 1000$. By studying the 
equations in detail one can see that, once the value of $\tilde
y_1/y_1$ is given, the neutrino sector of our model is uniquely 
fixed up to the two parameters $M_{\scr R} \sim M_{\scr I}$
and $z_2{ \tilde \omega}^{\scr (2)}_u$. We therefore look if there 
are solutions of the model for reasonable values of $\tilde y_1/y_1$,
$M_{\scr R}$ and $z_2 \tilde \omega^{\scr (2)}_u$ which predict
neutrino properties lying in the range allowed by oscillation 
experiments~\cite{nu1,nu2,nu,chooz}:
 \begin{eqnarray} \label{numcons1}
  |({\bf U})_{13}| \le 0.05 \hspace{3.72cm} & & \\
  0.49 \le |({\bf U})_{23}| \le 0.71 \hspace{3.72cm} & & \\ \nonumber
  0.03 \le |({\bf U})_{12}| \le 0.05 \quad 
  \textrm{(small angle MSW)} \hspace{0.07cm} & & \\
  \textrm{or} \; 0.35 \le |({\bf U})_{12}| \le 0.49 \quad 
  \textrm{(large angle MSW )} & & \\ \label{numcons2}
  50 \le \Delta m^2_{\scr \textrm{atm}}/\Delta m^2_{\scr \textrm{sun}} 
  \equiv (m^2_{\nu_3}-m^2_{\nu_2})/(m^2_{\nu_2}-m^2_{\nu_1})
  \le 1000 & &
\end{eqnarray}
The value of $y_1 M_{\scr R}$ will be fixed at the end to give the
exact absolute scale of the neutrino masses.

\begin{figure}
\begin{center}
\begingroup%
  \makeatletter%
  \newcommand{\GNUPLOTspecial}{%
    \@sanitize\catcode`\%=14\relax\special}%
  \setlength{\unitlength}{0.1bp}%
\begin{picture}(3600,2160)(0,0)%
\special{psfile=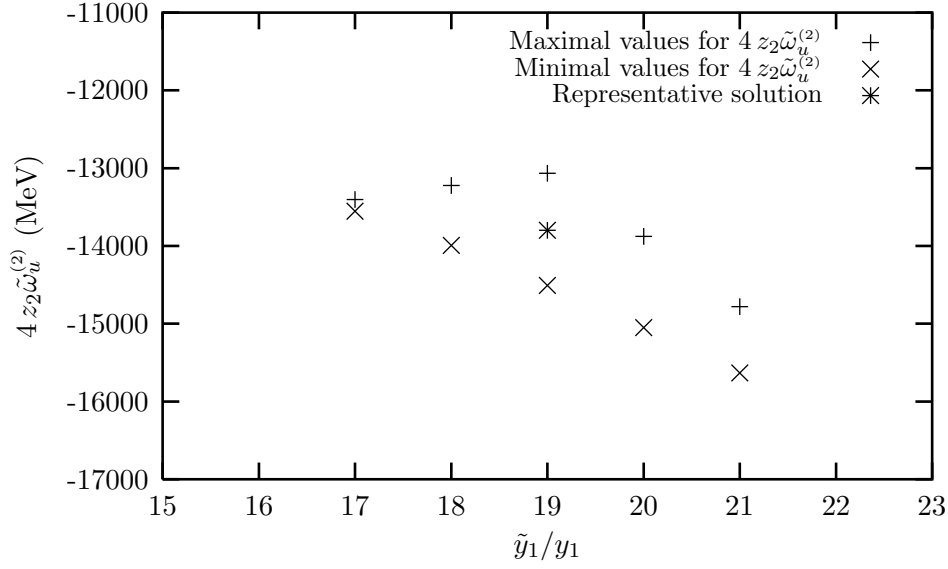 llx=0 lly=0 urx=720 ury=504 rwi=7200}
\put(3037,1747){\makebox(0,0)[r]{\small{Representative solution}}}%
\put(3037,1847){\makebox(0,0)[r]{\small{Minimal values for $4\,z_2\tilde\omega^{\scriptscriptstyle (2)}_u$}}}%
\put(3037,1947){\makebox(0,0)[r]{\small{Maximal values for $4\,z_2\tilde\omega^{\scriptscriptstyle (2)}_u$}}}%
\put(2000,50){\makebox(0,0){$\tilde y_1/y_1$}}%
\put(100,1180){%
\special{ps: gsave currentpoint currentpoint translate
270 rotate neg exch neg exch translate}%
\makebox(0,0)[b]{\shortstack{$4\,z_2\tilde\omega^{\scriptscriptstyle (2)}_u$ (MeV)}}%
\special{ps: currentpoint grestore moveto}%
}%
\put(3450,200){\makebox(0,0){23}}%
\put(3088,200){\makebox(0,0){22}}%
\put(2725,200){\makebox(0,0){21}}%
\put(2363,200){\makebox(0,0){20}}%
\put(2000,200){\makebox(0,0){19}}%
\put(1638,200){\makebox(0,0){18}}%
\put(1275,200){\makebox(0,0){17}}%
\put(913,200){\makebox(0,0){16}}%
\put(550,200){\makebox(0,0){15}}%
\put(500,2060){\makebox(0,0)[r]{-11000}}%
\put(500,1767){\makebox(0,0)[r]{-12000}}%
\put(500,1473){\makebox(0,0)[r]{-13000}}%
\put(500,1180){\makebox(0,0)[r]{-14000}}%
\put(500,887){\makebox(0,0)[r]{-15000}}%
\put(500,593){\makebox(0,0)[r]{-16000}}%
\put(500,300){\makebox(0,0)[r]{-17000}}%
\end{picture}%
\endgroup
 
\caption{Solution\,1 (large mixing MSW) in the 
$\tilde y_1/y_1$-$z_2\tilde \omega^{\scr (2)}_u$ parameter space \label{mod1}} 
\end{center}
\end{figure}

\begin{figure}
\begin{center}
\begingroup%
  \makeatletter%
  \newcommand{\GNUPLOTspecial}{%
    \@sanitize\catcode`\%=14\relax\special}%
  \setlength{\unitlength}{0.1bp}%
\begin{picture}(3600,2160)(0,0)%
\special{psfile=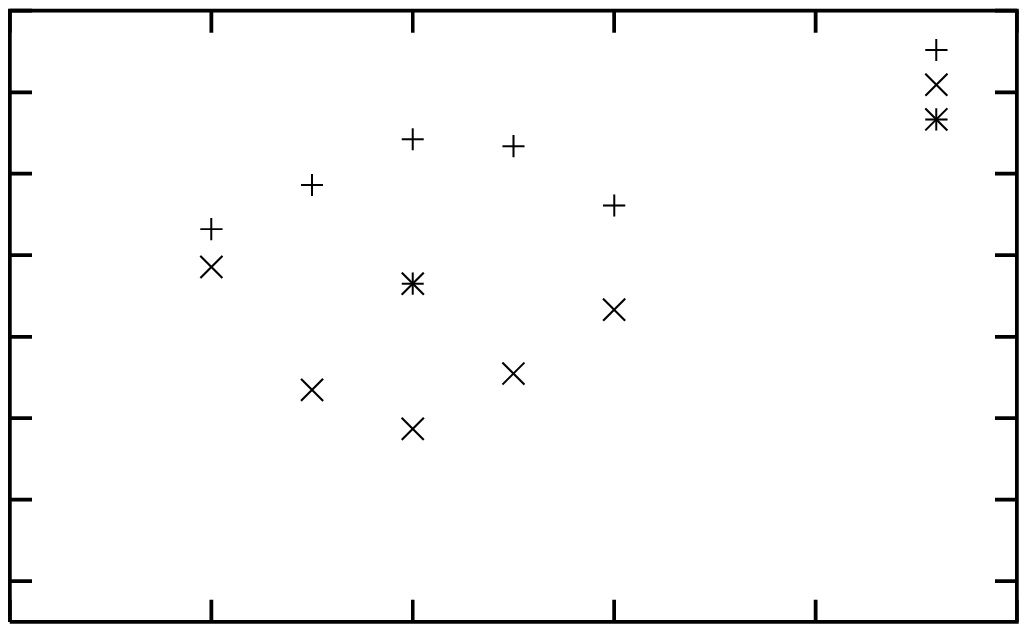 llx=0 lly=0 urx=720 ury=504 rwi=7200}
\put(3037,1747){\makebox(0,0)[r]{\small{Representative solution}}}%
\put(3037,1847){\makebox(0,0)[r]{\small{Minimal values for $4\,z_2\tilde\omega^{\scriptscriptstyle (2)}_u$}}}%
\put(3037,1947){\makebox(0,0)[r]{\small{Maximal values for $4\,z_2\tilde\omega^{\scriptscriptstyle (2)}_u$}}}%
\put(2000,50){\makebox(0,0){$\tilde y_1/y_1$}}%
\put(100,1180){%
\special{ps: gsave currentpoint currentpoint translate
270 rotate neg exch neg exch translate}%
\makebox(0,0)[b]{\shortstack{$4\,z_2\tilde\omega^{\scriptscriptstyle (2)}_u$ (MeV)}}%
\special{ps: currentpoint grestore moveto}%
}%
\put(3450,200){\makebox(0,0){30}}%
\put(2870,200){\makebox(0,0){28}}%
\put(2290,200){\makebox(0,0){26}}%
\put(1710,200){\makebox(0,0){24}}%
\put(1130,200){\makebox(0,0){22}}%
\put(550,200){\makebox(0,0){20}}%
\put(500,2060){\makebox(0,0)[r]{-30000}}%
\put(500,1825){\makebox(0,0)[r]{-32000}}%
\put(500,1591){\makebox(0,0)[r]{-34000}}%
\put(500,1356){\makebox(0,0)[r]{-36000}}%
\put(500,1121){\makebox(0,0)[r]{-38000}}%
\put(500,887){\makebox(0,0)[r]{-40000}}%
\put(500,652){\makebox(0,0)[r]{-42000}}%
\put(500,417){\makebox(0,0)[r]{-44000}}%
\end{picture}%
\endgroup
 
\caption{Solution\,2a (small mixing MSW) in the 
$\tilde y_1/y_1$-$z_2\tilde \omega^{\scr (2)}_u$ parameter space
 \label{mod2a}}
\end{center}
\end{figure}

\begin{figure}
\begin{center}
\begingroup%
  \makeatletter%
  \newcommand{\GNUPLOTspecial}{%
    \@sanitize\catcode`\%=14\relax\special}%
  \setlength{\unitlength}{0.1bp}%
\begin{picture}(3600,2160)(0,0)%
\special{psfile=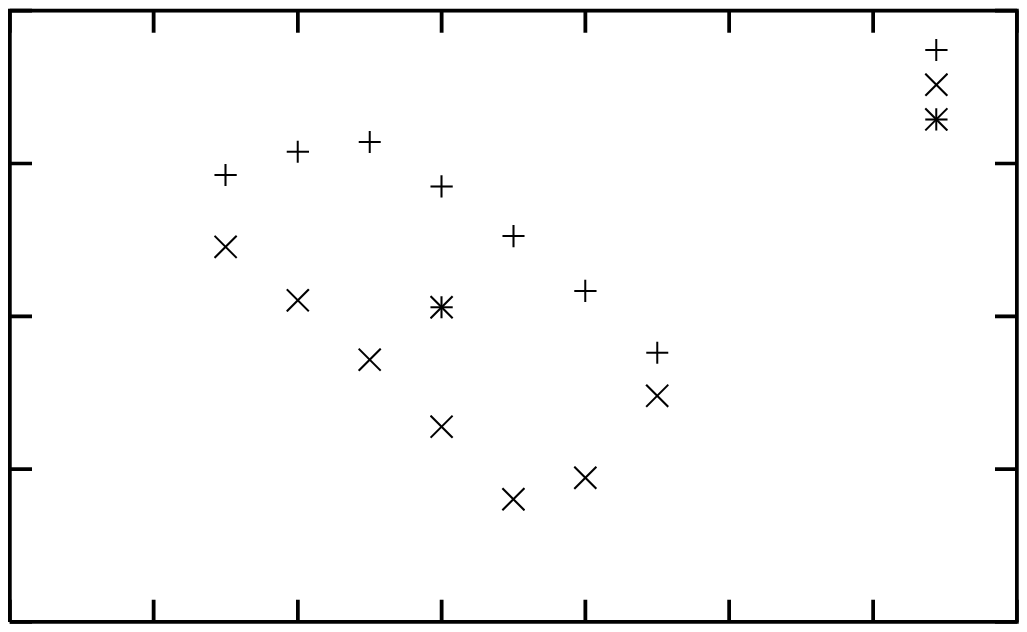 llx=0 lly=0 urx=720 ury=504 rwi=7200}
\put(3037,1747){\makebox(0,0)[r]{\small{Representative solution}}}%
\put(3037,1847){\makebox(0,0)[r]{\small{Minimal values for $4\,z_2\tilde\omega^{\scriptscriptstyle (2)}_u$}}}%
\put(3037,1947){\makebox(0,0)[r]{\small{Maximal values for $4\,z_2\tilde\omega^{\scriptscriptstyle (2)}_u$}}}%
\put(2000,50){\makebox(0,0){$\tilde y_1/y_1$}}%
\put(100,1180){%
\special{ps: gsave currentpoint currentpoint translate
270 rotate neg exch neg exch translate}%
\makebox(0,0)[b]{\shortstack{$4\,z_2\tilde\omega^{\scriptscriptstyle (2)}_u$ (MeV)}}%
\special{ps: currentpoint grestore moveto}%
}%
\put(3450,200){\makebox(0,0){-10}}%
\put(3036,200){\makebox(0,0){-12}}%
\put(2621,200){\makebox(0,0){-14}}%
\put(2207,200){\makebox(0,0){-16}}%
\put(1793,200){\makebox(0,0){-18}}%
\put(1379,200){\makebox(0,0){-20}}%
\put(964,200){\makebox(0,0){-22}}%
\put(550,200){\makebox(0,0){-24}}%
\put(500,2060){\makebox(0,0)[r]{-30000}}%
\put(500,1620){\makebox(0,0)[r]{-35000}}%
\put(500,1180){\makebox(0,0)[r]{-40000}}%
\put(500,740){\makebox(0,0)[r]{-45000}}%
\put(500,300){\makebox(0,0)[r]{-50000}}%
\end{picture}%
\endgroup
 
\caption{Solution\,2b (small mixing MSW) in the 
$\tilde y_1/y_1$-$z_2\tilde \omega^{\scr (2)}_u$ parameter space \label{mod2b}}
\end{center}
\end{figure}

Given these three parameters the model predicts the three neutrino
masses and three lepton mixing angles in ${\bf U}$.  We found two 
regions in the $\tilde y_1/y_1$-$z_2\tilde \omega^{\scr (2)}_u$
parameter space which obey the atmospheric neutrino requirements
together with small angle MSW~\cite{MSW} explanation for the solar 
neutrino puzzle (models\,2a,b) and one with the large angle MSW
(model\,1) as can be seen in the Figures \ref{mod1}, \ref{mod2a} and 
\ref{mod2b}.
For each region we fixed a representative solution (see Table~\ref{mmsols}) 
and used it to calculate the corresponding neutrino properties. 
The explicit results for the neutrinos are given in Table~\ref{ntrchar}.
\begin{table}
\begin{center}
\begin{tabular}{|l|l|r|c|}
\hline
Solution & MSW ef\/fect & $\tilde y_1/y_1$ & 
$z_2\tilde \omega^{\scr (2)}_u$ (MeV) \\
\hline \hline
Model\,1 & large mixing & $19$ & $-3450$ \\
Model\,2a & small mixing & $24$ & $-9175$ \\
Model\,2b &  small mixing & $-18$ & $-9925$ \\
\hline
\end{tabular}
\end{center}
\caption{Values of $\tilde y_1/y_1$ and $z_2 \tilde \omega^{\scr (2)}_u$ 
for the three representative solutions \label{mmsols}}
\end{table}
\begin{table}[h]
\begin{center}
\begin{tabular}{|c|r|r|r|}
\hline
Parameter & Value in Model\,1 & Value in Model\,2a & Value in Model\,2b \\
\hline \hline
$(y_1 M_{\scr R}/M_{\scr I}) \cdot m_{\nu_1}$ 
& $-0.0245$ eV & $-8.73 \cdot 10^{-4}$ eV & $1.16 \cdot 10^{-3}$ eV \\
$(y_1 M_{\scr R}/M_{\scr I}) \cdot m_{\nu_2}$ 
& $0.0876$ eV & $0.355 \,$ eV & $-0.467$ eV \\
$(y_1 M_{\scr R}/M_{\scr I}) \cdot m_{\nu_3}$ 
& $-2.402$ eV & $-3.031$ eV & $4.365$ eV \\
\hline
$\theta^{\scr (\nu)}_{\scr 12}$ & $-0.487$ & $-0.050$ & $0.051$ \\
$\theta^{\scr (\nu)}_{\scr 23}$ & $0.205$ & $0.506$ & $0.496$ \\
$\theta^{\scr (\nu)}_{\scr 31}$ & $0.004$ & $0.003$ & $-0.003$ \\
\hline
$m_{\nu_2}/m_{\nu_1}$ & $-3.57$ & $-406.3$ & $-401.2$ \\
$m_{\nu_3}/m_{\nu_2}$ & $-27.43$ & $-8.54$ & $-9.35$ \\
$\left( \dfrac{m_{\nu_3}^2-m_{\nu_2}^2}{m_{\nu_2}^2-m_{\nu_1}^2} \right)$ 
& $815.4$ & $71.9$ & $86.4$ \\
\hline
\end{tabular}
\end{center}
\caption{\label{ntrchar} Masses and mixing angles of the light 
  neutrinos at $M_{\scr Z}$}
\end{table}

In this way the explicit LH and RH mixing matrices are also
fixed. The corresponding mixing angles are given in Table~\ref{mixang}.
They are used to calculate the branching ratios of the nucleon decays. 
\begin{table}[h]
\begin{center}
\begin{tabular}{|c|r|r|r|}
\hline
Parameter & Value in model\,1 & Value in model\,2a & Value in model\,2b \\
\hline \hline
$\theta^{\scr (u)}_{\scr L12}$ & $-0.627$ & $-0.540$ & $0.275$ \\
$\theta^{\scr (u)}_{\scr L23}$ & $-0.055$ & $-0.056$ & $-0.075$ \\
$\theta^{\scr (u)}_{\scr L31}$ & $0.000$ & $0.000$ & $0.000$ \\
\hline
$\theta^{\scr (u)}_{\scr R12}$ & $0.005$ & $0.006$ & $-0.012$ \\
$\theta^{\scr (u)}_{\scr R23}$ & $0.049$ & $0.051$ & $0.043$ \\
$\theta^{\scr (u)}_{\scr R31}$ & $0.000$ & $0.000$ & $0.000$ \\
\hline \hline
$\theta^{\scr (d)}_{\scr L12}$ & $-0.404$ & $-0.317$ & $0.498$ \\
$\theta^{\scr (d)}_{\scr L23}$ & $-0.025$ & $-0.024$ & $-0.041$ \\
$\theta^{\scr (d)}_{\scr L31}$ & $-0.018$ & $-0.016$ & $0.012$ \\
\hline
$\theta^{\scr (d)}_{\scr R12}$ & $0.117$ & $0.152$ & $-0.092$ \\
$\theta^{\scr (d)}_{\scr R23}$ & $0.979$ & $1.011$ & $0.679$ \\
$\theta^{\scr (d)}_{\scr R31}$ & $0.000$ & $0.000$ & $0.000$ \\
\hline \hline
$\theta^{\scr (e)}_{\scr L12}$ & $-0.018$ & $-0.015$ & $0.015$ \\
$\theta^{\scr (e)}_{\scr L23}$ & $0.733$ & $-0.060$ & $-0.089$ \\
$\theta^{\scr (e)}_{\scr L31}$ & $0.000$ & $-0.001$ & $0.001$ \\
\hline
$\theta^{\scr (e)}_{\scr R12}$ & $0.262$ & $0.314$ & $-0.301$ \\
$\theta^{\scr (e)}_{\scr R23}$ & $-0.063$ & $0.747$ & $0.561$ \\
$\theta^{\scr (e)}_{\scr R31}$ & $0.014$ & $-0.001$ & $0.002$ \\
\hline \hline
$\theta^{\scr (\nu)}_{\scr 12}$ & $-0.487$ & $-0.050$ & $0.051$ \\
$\theta^{\scr (\nu)}_{\scr 23}$ & $0.205$ & $0.506$ & $0.496$ \\
$\theta^{\scr (\nu)}_{\scr 31}$ & $0.004$ & $0.003$ & $-0.003$ \\
\hline
\end{tabular}
\caption{Mixing angles in the three different models \label{mixang}}
\end{center}
\end{table}

\section{Calculation of the nucleon decay rates}

The partial decay rate for a given process nucleon $\rightarrow$ 
meson + antilepton is expressed as follows:
\begin{equation} \label{decform}
\Gamma^{}_j \; = \; \dfrac{1}{16\pi} \, m^2_{\scr \textrm{nucl}} \,
\rho_{\scr j} \, |S|^2 \, |\mathcal{A}_{}|^2 \,
\Big( |\mathcal{A}^{}_{\scr L}|^2 \sum_l |A^{}_{l} \mathcal{M}^{}_{l}|^2
+ |\mathcal{A}^{}_{\scr R}|^2 \sum_r |A^{}_{r} \mathcal{M}^{}_{r}|^2 
\Big) \; ,
\end{equation}
where $\mathcal{M}^{}_l$ and $\mathcal{M}^{}_r$ are the hadronic 
transition matrix elements for the relevant decay process. $l$ and $r$ 
denote the chirality of the corresponding antilepton. $A^{}_{l}$ 
and $A^{}_{r}$ are the relevant coefficients of the effective 
Lagrangian (\ref{eldfnd}) given in App.\,II. $\mathcal{A}_{}$, 
$\mathcal{A}^{}_{\scr L}$ and $\mathcal{A}^{}_{\scr R}$ are factors 
which result from the renormalization of the four fermion operators 
as follows:
\begin{eqnarray}
\mathcal{A}^{}_{\scr L} & = & \Big( 
\dfrac{\alpha^{}_{\scr 1}(M^{}_{\scr Z})}{\alpha^{}_{\scr 1}(M^{}_{\scr I})} 
\Big)_{}^{-\frac{23}{82}} \\
\mathcal{A}^{}_{\scr R} & = & \Big( 
\dfrac{\alpha^{}_{\scr 1}(M^{}_{\scr Z})}{\alpha^{}_{\scr 1}(M^{}_{\scr I})} 
\Big)_{}^{-\frac{11}{82}} \\ \nonumber
\mathcal{A} \hspace{0.2cm} & = & \Big( 
\dfrac{\alpha^{}_{\scr 4C}(M^{}_{\scr I})}{\alpha^{}_{\scr 4C}(M^{}_{\scr U})} 
\Big)_{}^{-\frac{5}{8}}
\Big( \dfrac{\alpha^{}_{\scr 2L}(M^{}_{\scr I})}{\alpha^{}_{\scr 2L}
(M^{}_{\scr U})} \Big)_{}^{-\frac{27}{100}}
\Big( \dfrac{\alpha^{}_{\scr 2R}(M^{}_{\scr I})}{\alpha^{}_{\scr 2R}
(M^{}_{\scr U})} \Big)_{}^{-\frac{3}{20}} 
\Big( \dfrac{\alpha^{}_{\scr 2}(M^{}_{\scr Z})}{\alpha^{}_{\scr 2}
(M^{}_{\scr I})} \Big)_{}^{\frac{27}{38}}
\\ & &  \hspace{-0.16cm} \cdot \,
\Big( \dfrac{\alpha^{}_{\scr 3}(M^{}_{\scr Z})}{\alpha^{}_{\scr 3}
(M^{}_{\scr I})} \Big)_{}^{\frac{2}{7}}
\Big( \dfrac{\alpha^{}_{\scr 3}(m^{}_b)}
{\alpha^{}_{\scr 3}(M^{}_{\scr Z})} \Big)_{}^{\frac{6}{23}}
\Big( \dfrac{\alpha^{}_{\scr 3}(m^{}_c)}
{\alpha^{}_{\scr 3}(m^{}_b)} \Big)_{}^{\frac{6}{25}}
\Big( \dfrac{\alpha^{}_{\scr 3}(1 \; {\textrm{GeV}})}
{\alpha^{}_{\scr 3}(m^{}_c)} \Big)_{}^{\frac{2}{9}}
\end{eqnarray}
Using then \cite{buras}
\begin{equation}
\alpha^{}_{\scr 3}(1 \; {\textrm{GeV}}) \; = \; 0.544 \; , \quad
\alpha^{}_{\scr 3}(m^{}_c) \; = \; 0.412 \; , \quad
\alpha^{}_{\scr 3}(m^{}_b) \; = \; 0.226
\end{equation}
one obtains
\begin{equation}|\mathcal{A}^{}_{\scr L}|^2 \; = \; 1.155 \; , \quad
|\mathcal{A}^{}_{\scr R}|^2 \; = \; 1.071 \; , \quad
|\mathcal{A}|^2 \; = \; 23.59 \, .
\end{equation}
$|S|_{}^2 = \langle \Psi^s_{\scr \textrm{Nucl}} 
(\vec r^{}_{\scr 1},\vec r^{}_{\scr 2},\vec r^{}_{\scr 3}) \, | \, 
\delta(\vec r^{}_{\scr 1}-\vec r^{}_{\scr 2}) \, | \,
\Psi^s_{\scr \textrm{Nucl}} 
(\vec r^{}_{\scr 1},\vec r^{}_{\scr 2},\vec r^{}_{\scr 3}) \rangle$  
is the probability to find two valence quarks of the nucleon at 
one point in space. We used here the value $|S|_{}^2 = 0.012$ GeV$^3$
given in~\cite{kr}. $\rho_{\scr j} \equiv (1-\chi_j^2)(1-\chi_j^4)$ 
with $\chi_j^{}=m^{}_{\scr \textrm{Meson}}/m^{}_{\scr \textrm{Nucl}}$ 
is an $SU(6)$ spin-flavour symmetry breaking phase space factor.

The resulting branching ratios are given in the Tables \ref{pdrt}
and \ref{ndrt}. Our model also gives the total decay rates. The 
predicted rates in the different models are given in the 
Tables~\ref{tdrpt} and \ref{tdrnt}. Note however that these
predictions have a relatively large uncertainty. This was estimated 
by Langacker~\cite{lang} to be
\begin{equation}
\Delta\tau_{p\rightarrow e^+ \pi^{\circ}} = 10^{\pm 0.7 \pm {1.0 
^{+0.5}_{-3.0}}} \; \textrm{yrs} .
\end{equation}
Taking this into account our predicted representative rates have therefore a 
chance to be observed by SuperKamiokande~\cite{SK} and ICARUS~\cite{ICAR}.

Yet the essential predictions of the model are actually the branching 
ratios. Comparing our branching ratios with those of the conventional 
$SO(10)$ (i.e. without large mixings) one sees clearly the suppression 
of the $p,n \rightarrow e^+ X $ channels relative to the channels 
$p,n \rightarrow {\mu}^+ X , {\nu}^c X$ . In particular 
$p,n \rightarrow {\mu}^+ \pi , {\nu}^c K$ are prominant. Note that in
SUSY GUTs~\cite{sgut} the dominant decays are into final states 
involving $K$ mesons~\footnote{This is also the case for non-SUSY 
GUTs with maximal RH mixings~\cite{plr}.}. The special properties 
of our model are clearly reflected in the comparison of ratios as is 
done in Table~\ref{rocdr}.

\section{Conclusions}

We presented an $SO(10)$ GUT with a global $U(1)_{\scr F}$ family
symmetry that is consistent with all experimental observations. The 
special thing about this model is that it involves large mixing angles 
for the quarks, in contrast with the conventional expectations. The large 
mixing in the lepton (neutrino) sector results therefore naturally.

Large rotations have considerable consequences for observables outside 
the Standard Model. In particular calculations of the nucleon decay 
rates require the knowledge of all mixing angles. We found rates 
which are obviously different from the conventional GUTs.
Our model is however only one simple example for the effects of 
large mixing angles. We are studying now a SUSY GUT with all possible 
phases and its consequences also for other GUT observables or RH
currents. The hope is that by restricting the values of the mixing 
angles, we will be able to reduce the large freedom in the present 
models for fermionic masses.

\subsection*{Acknowledgments}

We would like to thank M.\ K.\ Parida for discussions and especially 
for helping us with threshold effects. One of us (Y.A) is also grateful
to the School of Physics and Astronomy of the Tel Aviv University for
the kind hospitality during his present visit.

\begin{table}
\begin{center}
\begin{tabular}{|l|r|r|r|r|}
\hline
Decay channel & Rates in \% & Rates in \% & Rates in \% & Rates in \% \\
of the proton & (no mixing) & in model\,1 & in model\,2a & in model\,2b \\
\hline \hline
$p \; \rightarrow \; \epi$  
& 33.6 \; & 21.4 \; & 25.1 \; & 27.8 \; \\
$p \; \rightarrow \; \ek$   
&  --- \; &  3.1 \; &  2.6 \; &  4.5 \; \\
$p \; \rightarrow \; \et$   
&  1.2 \; &  0.8 \; &  0.9 \; &  1.0 \; \\
$p \; \rightarrow \; \mpi$  
&  --- \; &  8.5 \; &  5.7 \; &  5.6 \; \\
$p \; \rightarrow \; \muk$  
&  5.8 \; &  2.6 \; &  0.9 \; &  1.8 \; \\
$p \; \rightarrow \; \mt$   
&  --- \; &  0.3 \; &  0.2 \; &  0.2 \; \\
$p \; \rightarrow \; \ero$ 
&  5.1 \; &  3.3 \; &  3.8 \; &  4.2 \; \\
$p \; \rightarrow \; \eo$   
& 16.9 \; & 10.8 \; & 12.7 \; & 14.0 \; \\
$p \; \rightarrow \; \eks$  
&  --- \; &  0.0 \; &  0.0 \; &  0.0 \; \\
$p \; \rightarrow \; \mro$  
&  --- \; &  1.3 \; &  0.9 \; &  0.8 \; \\
$p \; \rightarrow \; \mo$   
&  --- \; &  4.3 \; &  2.9 \; &  2.8 \; \\
$p \; \rightarrow \; \nep$  
& 32.3 \; & 25.6 \; & 35.6 \; & 27.7 \; \\
$p \; \rightarrow \; \nek$  
&  --- \; &  2.0 \; &  2.0 \; &  4.1 \; \\
$p \; \rightarrow \; \nmpi$ 
&  --- \; &  8.9 \; &  0.5 \; &  0.3 \; \\
$p \; \rightarrow \; \nmk$  
&  0.1 \; &  1.3 \; &  0.2 \; &  0.3 \; \\
$p \; \rightarrow \; \nero$ 
&  4.9 \; &  3.9 \; &  5.4 \; &  4.2 \; \\
$p \; \rightarrow \; \neks$ 
&  --- \; &  0.4 \; &  0.3 \; &  0.6 \; \\
$p \; \rightarrow \; \nmro$ 
&  --- \; &  1.4 \; &  0.1 \; &  0.0 \; \\
$p \; \rightarrow \; \nmks$ 
&  0.1 \; &  0.0 \; &  0.0 \; &  0.0 \; \\
$p \; \rightarrow \; \ntp$  
&  --- \; &  0.1 \; &  0.1 \; &  0.0 \; \\
$p \; \rightarrow \; \ntk$  
&  --- \; &  0.1 \; &  0.1 \; &  0.0 \; \\
$p \; \rightarrow \; \ntro$ 
&  --- \; &  0.0 \; &  0.0 \; &  0.0 \; \\
$p \; \rightarrow \; \ntks$ 
&  --- \; &  0.0 \; &  0.0 \; &  0.0 \; \\
\hline
$p \; \rightarrow \; e^{\scr +} X^{\scr 0}$ 
& 56.8 \; & 39.4 \; & 45.1 \; & 51.5 \; \\
$p \; \rightarrow \; \mu^{\scr +} X^{\scr 0}$ 
&  5.8 \; & 17.0 \; & 10.6 \; & 11.2 \; \\
$p \; \rightarrow \; \nu^{\scr C} X^{\scr +}$ 
& 37.4 \; & 43.7 \; & 44.3 \; & 37.2 \; \\
\hline
\end{tabular}
\caption{Partial decay rates $\Gamma_i/\Gamma$ of the proton for
the case of vanishing fermionic mixing in comparison with the rates obtained
in our solutions \label{pdrt}}
\end{center}
\end{table}
\begin{table}
\begin{center}
\begin{tabular}{|l|c|c|c|}
\hline
Quantity & Value in model\,1 & Value in model\,2a & Value in model\,2b \\
\hline \hline
$\Gamma_p$ & $2.54 \cdot 10_{}^{\scr -35} \; \textrm{yr}_{}^{\scr -1}$
& $2.57 \cdot 10_{}^{\scr -35} \; \textrm{yr}_{}^{\scr -1}$
& $2.83 \cdot 10_{}^{\scr -35} \; \textrm{yr}_{}^{\scr -1}$ \\
$\tau_p$ & $3.94 \cdot 10_{}^{\scr 34} \; \textrm{yr}$ 
& $3.89 \cdot 10_{}^{\scr 34} \; \textrm{yr}$ 
& $3.53 \cdot 10_{}^{\scr 34} \; \textrm{yr}$ \\
\hline
\end{tabular}
\caption{Total decay rate and lifetime of the proton
for the representative solutions \label{tdrpt}}
\end{center}
\end{table}
\begin{table}
\begin{center}
\begin{tabular}{|l|r|r|r|r|}
\hline
Decay channel & Rates in \% & Rates in \% & Rates in \% & Rates in \% \\
of the neutron & (no mixing) & in model\,1 & in model\,2a 
& in model\,2b \\
\hline \hline
$n \; \rightarrow \; \epin$  
& 62.9 \; & 40.1 \; & 46.2 \; & 49.9 \; \\
$n \; \rightarrow \; \mpin$  
&  --- \; & 15.8 \; & 10.4 \; & 10.0 \; \\
$n \; \rightarrow \; \eron$  
&  9.7 \; &  6.2 \; &  7.1 \; &  7.7 \; \\
$n \; \rightarrow \; \mron$  
&  --- \; &  2.4 \; &  1.6 \; &  1.5 \; \\
$n \; \rightarrow \; \nepn$  
& 15.1 \; & 12.0 \; & 16.4 \; & 12.5 \; \\
$n \; \rightarrow \; \nekn$  
&  --- \; &  6.8 \; &  4.9 \; &  8.2 \; \\
$n \; \rightarrow \; \netn$  
&  0.6 \; &  0.4 \; &  0.6 \; &  0.5 \; \\
$n \; \rightarrow \; \nmpin$ 
&  --- \; &  4.2 \; &  0.2 \; &  0.1 \; \\
$n \; \rightarrow \; \nmkn$  
&  1.7 \; &  0.1 \; &  1.0 \; &  0.9 \; \\
$n \; \rightarrow \; \nmtn$  
&  --- \; &  0.2 \; &  0.0 \; &  0.0 \; \\
$n \; \rightarrow \; \neron$ 
&  2.3 \; &  1.8 \; &  2.5 \; &  1.9 \; \\
$n \; \rightarrow \; \neon$  
&  7.7 \; &  6.1 \; &  8.4 \; &  6.4 \; \\
$n \; \rightarrow \; \neksn$ 
&  --- \; &  0.2 \; &  0.2 \; &  0.4 \; \\
$n \; \rightarrow \; \nmron$ 
&  --- \; &  0.6 \; &  0.0 \; &  0.0 \; \\
$n \; \rightarrow \; \nmon$  
&  --- \; &  2.1 \; &  0.1 \; &  0.1 \; \\
$n \; \rightarrow \; \nmksn$ 
&  0.0 \; &  0.1 \; &  0.0 \; &  0.0 \; \\
$n \; \rightarrow \; \ntpn$  
&  --- \; &  0.0 \; &  0.0 \; &  0.0 \; \\
$n \; \rightarrow \; \ntkn$  
&  --- \; &  0.7 \; &  0.4 \; &  0.0 \; \\
$n \; \rightarrow \; \nttn$  
&  --- \; &  0.0 \; &  0.0 \; &  0.0 \; \\
$n \; \rightarrow \; \ntron$ 
&  --- \; &  0.0 \; &  0.0 \; &  0.0 \; \\
$n \; \rightarrow \; \nton$  
&  --- \; &  0.0 \; &  0.0 \; &  0.0 \; \\
$n \; \rightarrow \; \ntksn$ 
&  --- \; &  0.0 \; &  0.0 \; &  0.0 \; \\
\hline
$n \; \rightarrow \; e^{\scr +} X^{\scr -}$ 
& 72.6 \; & 46.3 \; & 53.3 \; & 57.6 \; \\
$n \; \rightarrow \; \mu^{\scr +} X^{\scr -}$ 
&  --- \; & 18.2 \; & 12.0 \; & 11.5 \; \\
$n \; \rightarrow \; \nu^{\scr C} X^{\scr 0}$ 
& 27.4 \; & 35.3 \; & 34.7 \; & 31.0 \; \\
\hline
\end{tabular}
\caption{Partial decay rates $\Gamma_i/\Gamma$ of the bound neutron for
the case of vanishing fermionic mixing in comparison with the rates 
obtained in our solutions \label{ndrt}}
\end{center}
\end{table}
\begin{table}
\begin{center}
\begin{tabular}{|l|c|c|c|}
\hline
Quantity & Value in model\,1 & Value in model\,2a & Value in model\,2b \\
\hline \hline
$\Gamma_n$ & $2.72 \cdot 10_{}^{\scr -35} \; \textrm{yr}_{}^{\scr -1}$
& $2.80 \cdot 10_{}^{\scr -35} \; \textrm{yr}_{}^{\scr -1}$
& $3.14 \cdot 10_{}^{\scr -35} \; \textrm{yr}_{}^{\scr -1}$ \\
$\tau_n$ & $3.68 \cdot 10_{}^{\scr 34} \; \textrm{yr}$ 
& $3.57 \cdot 10_{}^{\scr 34} \; \textrm{yr}$ 
& $3.18 \cdot 10_{}^{\scr 34} \; \textrm{yr}$ \\
\hline
\end{tabular}
\caption{Total decay rate and lifetime of the bound neutron 
for the representative solutions \label{tdrnt}}
\end{center}
\end{table}
\begin{table}
\begin{center}
\begin{tabular}{|c|c|c|c|c|}
\hline
Ratio & No mixing & \; Model\,1 \; & \; Model\,2a \; & \; Model\,2b \; \\
\hline \hline
$\dfrac{\Gamma(p \rightarrow \ek)}{\Gamma(p \rightarrow \epi)}$ 
& 0 & 0.145 & 0.104 & 0.162 \\ \hline
$\dfrac{\Gamma(p \rightarrow \mpi)}{\Gamma(p \rightarrow \muk)}$ 
& 0 & 3.27 \; & 6.33 \; & 3.11 \; \\ \hline
$\dfrac{\Gamma(p \rightarrow \nu^{\scr C}_{} K^+)}
{\Gamma(p \rightarrow \nu^{\scr C}_{} \pi^+)}$ 
& 0.003 & 0.098 & 0.064 & 0.157 \\ \hline
$\dfrac{\Gamma(p \rightarrow \epi)}
{\Gamma(p \rightarrow \nu^{\scr C}_{} \pi^+)}$ 
& 1.040 & 0.618 & 0.693 & 0.993 \\
\hline \hline
$\dfrac{\Gamma(n \rightarrow \mpin)}{\Gamma(n \rightarrow \epin)}$ 
& 0 & 0.394 & 0.225 & 0.200 \\ \hline
$\dfrac{\Gamma(n \rightarrow \mron)}{\Gamma(n \rightarrow \eron)}$ 
& 0 & 0.387 & 0.225 & 0.195 \\ \hline
$\dfrac{\Gamma(n \rightarrow \nu^{\scr C}_{} K^0)}
{\Gamma(n \rightarrow \nu^{\scr C}_{} \pi^0)}$ 
& 0.113 & 0.469 & 0.347 & 0.722 \\ \hline
$\dfrac{\Gamma(n \rightarrow \epin)}
{\Gamma(n \rightarrow \nu^{\scr C}_{} \pi^0)}$ 
& 4.16 \; & 2.48 \; & 2.78 \; & 3.96 \; \\
\hline
\end{tabular}
\end{center}
\caption{Ratios of some partial nucleon decay rates 
\label{rocdr}}
\end{table}

\newpage

\section*{Appendix\,I: Fermionic masses in $SO(10)$ GUTs}

Dirac masses for the charged fermions are of the form 
$m ( \Psi^{\scr C}_{\scr L} )^T C \Psi^{}_{\scr L}$
and therefore transform under $SO(10)$ as follows:
\begin{equation}
{\bf 16} \otimes {\bf 16} \; = \; ({\bf 10} \oplus {\bf 126})_{\textrm{symm}}
\oplus {\bf 120}_{\textrm{antisymm}}
\end{equation}
The $SO(10)$ Higgs representations which can give masses to the fermions 
have the representation content
\begin{eqnarray}
{\bf 10}  & \longrightarrow & ({\bf 1,2,2}) \oplus ({\bf 6,1,1}) \\
{\bf 120} & \longrightarrow & ({\bf 1,2,2}) \oplus ({\bf 15,2,2})
\oplus ({\bf 6,3,1}) \oplus ({\bf 6,1,3}) \oplus ({\bf 10,1,1}) 
\oplus ({\bf \overline{10},1,1}) \\
{\bf 126} & \longrightarrow & ({\bf 15,2,2}) \oplus ({\bf 10,3,1}) 
\oplus ({\bf \overline{10},1,3}) \oplus ({\bf 6,1,1})
\end{eqnarray}
under $G_{\scr \textrm{PS}}= SU(4)_{\scr C} \otimes SU(2)_{\scr L} 
\otimes SU(2)_{\scr R}$, while for 
the fermions in the ${\bf 16}$
\begin{eqnarray}
{\bf 16} & \longrightarrow & ({\bf 4,2,1}) \oplus ({\bf {\overline 4},1,2}) 
\end{eqnarray}
Hence the Dirac mass terms transform as follows
\begin{equation}
({\bf 4,2,1}) \otimes ({\bf \bar 4,1,2}) = ({\bf 15,2,2}) 
\oplus ({\bf 1,2,2})
\end{equation}
The different Yukawa couplings and the corresponding VEVs of 
$\Phi_{\bf 10}$, $\Phi_{\bf 120}$ and $\Phi_{\bf 126}$
are given in Table~\ref{hgs}.
\begin{table}[h]
\begin{center}
\begin{tabular}{|l||c|c|c|c|}
\hline
Higgs representation & $({\bf 1,2,2})_{\bf \scr 10}^{\scr (i)}$ & 
$({\bf 1,2,2})_{\bf \scr 120}^{\scr (j)}$ & 
$({\bf 15,2,2})_{\bf \scr 120}^{\scr (k)}$ & 
$({\bf 15,2,2})_{\bf \scr 126}^{\scr (l)}$ \\ \hline
Yukawa coupling matrix & ${\bf Y}_{\bf \scr 10}^{\scr (i)}$ & 
${\bf Y}_{\bf \scr 120}^{\scr (j)}$ & ${\bf Y}_{\bf \scr 120}^{\scr (k)}$ & 
${\bf Y}_{\bf \scr 126}^{\scr (l)}$ \\ \hline
Vacuum expectation values & $\upsilon^{\scr (i)}_u$, $\upsilon^{\scr (i)}_d$ & 
$\tilde \upsilon^{\scr (j)}_u$, $\tilde \upsilon^{\scr (j)}_d$ & 
$\tilde \omega^{\scr (k)}_u$, $\tilde \omega^{\scr (k)}_d$ & 
$\omega^{\scr (l)}_u$, $\omega^{\scr (l)}_d$ \\ \hline
\end{tabular}
\end{center}
\caption{Higgs couplings and vacuum expectation values in $SO(10)$ GUTs 
\label{hgs}}
\end{table}

Taking now the Clebsch-Gordan coefficients into account the mass
matrices get the following contributions~\cite{gena}
\begin{eqnarray} \label{msrel1}
{\bf M}^{}_d \hspace{0.55cm} & = & 
\upsilon_d {\bf Y_{\scr 10}} + \omega_d {\bf Y_{\scr 126}}
+ (\tilde \upsilon_d + \tilde \omega_d) {\bf Y_{\scr 120}} \\
{\bf M}^{}_e \hspace{0.55cm} & = & 
\upsilon_d {\bf Y_{\scr 10}} - 3 \, \omega_d {\bf Y_{\scr 126}}
+ (\tilde \upsilon_d - 3 \, \tilde \omega_d) {\bf Y_{\scr 120}} \\
{\bf M}^{}_u \hspace{0.5cm} & = & 
\upsilon_u {\bf Y_{\scr 10}} + \omega_u {\bf Y_{\scr 126}}
+ (\tilde \upsilon_u + \tilde \omega_u) {\bf Y_{\scr 120}} \\ \label{msrel2a}
{\bf M}^{\scr (\textrm{Dir})}_\nu \hspace{0.1cm} & = & 
\upsilon_u {\bf Y_{\scr 10}} - 3 \, \omega_u {\bf Y_{\scr 126}}
+ (\tilde \upsilon_u - 3 \, \tilde \omega_u) {\bf Y_{\scr 120}}
\end{eqnarray}
The RH neutrinos can acquire also Majorana masses in term of the 
$({\bf 10,1,3})$ component of $\Phi_{\bf 126}$:
\begin{eqnarray}  \label{majmas2}
({\bf \bar 4,1,2}) \otimes ({\bf \bar 4,1,2}) \otimes ({\bf 10,1,3})
& = & ({\bf 1,1,1}) \oplus \; \dots \dots 
\end{eqnarray}
\begin{equation}
{\bf M}^{\scr (\textrm{Maj})}_{\nu \scr{R}} 
\; = \; M_{\scr R} {\bf Y_{\scr 126}} \; \sim \; 
M_{\scr I} {\bf Y_{\scr 126}}
\end{equation}
The corresponding VEV $M_{\scr I}$ is responsible for the symmetry
breaking step $G_{\scr \textrm{PS}} \rightarrow G_{\scr \textrm{SM}}$.

The neutrinos will have therefore the $6 \times 6$ mass martix
\begin{equation}
{\bf M} \; = \; \begin{pmatrix}
{\bf 0} & 
{\bf M}^{\scr (\textrm{Dir})}_\nu \\
\big( {\bf M}^{\scr (\textrm{Dir})}_\nu \big)^{\scr T}
& {\bf M}^{\scr (\textrm{Maj})}_{\nu \scr{R}} 
\end{pmatrix}
\end{equation}
Using the fact that the non-vanishing entries of the Majorana mass
matrix are much larger than those of the Dirac matrix one can
approximately block diagonalize the  $6 \times 6$ matrix and obtain
the see-saw matrix~\cite{ss} 
\begin{equation} \label{seesmass}
{\bf M}_\nu^{\scr \textrm{light}}
\; \approx \; - {\bf M}^{\scr (\textrm{Dir})}_\nu 
\big( {\bf M}^{\scr (\textrm{Maj})}_{\nu \scr R} \big)^{-1}
\big( {\bf M}^{\scr (\textrm{Dir})}_\nu \big)^T
\end{equation}
as well as 
\begin{equation}
{\bf M}_\nu^{\scr \textrm{heavy}}\; \approx \; 
{\bf M}^{\scr (\textrm{Maj})}_{\nu \scr R}.
\end{equation}
${\bf M}_\nu^{\scr \textrm{light}}$ is symmetric and therefore can be
diagonalized using one unitary matrix ${\bf N}_\nu$
\begin{equation}
{\bf N}_\nu^T {\bf M}_\nu^{\scr \textrm{light}} {\bf N}_\nu \; = \; 
{\bf M}_\nu^{\scr \textrm{light(D)}}
\end{equation}
Neutrino oscillations are induced via the leptonic analogue to the CKM matrix
\begin{equation}
{\bf U} = {\bf E}_{\scr L}^{\dagger} {\bf N}_\nu
\end{equation}

\section*{Appendix II: Effective $SO(10)$ Lagrangian for nucleon decays}

The baryon number violating part of the $SO(10)$ Lagrangian (without
fermionic mixing) is known to be~\cite{gut}
\begin{eqnarray} \nonumber
  \mathcal{L}_{\scr \Delta B \ne 0}
& = & \dfrac{g^{}_{\scr U}}{\sqrt{2}} \; \bar 
  X^{\scr \alpha}_{\scr \mu} \;
  \big( \varepsilon_{\scr \alpha \beta \gamma} \bar u_{\scr L}^{\scr C
    \gamma}  \gamma^{\scr \mu} u_{\scr L}^{\scr \beta} + \bar d_{\scr L
  \alpha}^{} \gamma^{\scr \mu} e^{\scr +}_{\scr L} 
   + \bar d_{\scr R \alpha}^{} \gamma^{\scr \mu} e^{\scr +}_{\scr R}
  \big) \\ \nonumber
& + & \dfrac{g^{}_{\scr U}}{\sqrt{2}} \; 
  \bar Y^{\scr \alpha}_{\scr \mu} \; \big( 
  \varepsilon_{\scr \alpha \beta \gamma} \bar u_{\scr L}^{\scr C \gamma} 
  \gamma^{\scr \mu} d_{\scr L}^{\scr \beta}
  - \bar d_{\scr R \alpha}^{}  \gamma^{\scr \mu} \nu_{\scr e R}^{\scr C}
  - \bar u_{\scr L \alpha}^{} \gamma^{\scr \mu} e^{\scr +}_{\scr L}
  \big) \\ \nonumber
& + & \dfrac{g^{}_{\scr U}}{\sqrt{2}} \; X^{' \scr \alpha}_{\scr \mu} \; \big( 
- \varepsilon_{\scr \alpha \beta \gamma} \bar d_{\scr L}^{\scr C \gamma}
  \gamma^{\scr \mu} d_{\scr L}^{\scr \beta}
  - \bar u_{\scr L \alpha}^{} \gamma^{\scr \mu} \nu^{\scr C}_{\scr e L} 
  - \bar u_{\scr R \alpha}^{} \gamma^{\scr \mu} \nu^{\scr C}_{\scr e R}
  \big) \\ \nonumber
& + & \dfrac{g^{}_{\scr U}}{\sqrt{2}} \; Y^{' \scr \alpha}_{\scr \mu} \; \big( 
  \varepsilon_{\scr \alpha \beta \gamma} \bar d_{\scr L}^{\scr C \gamma}
  \gamma^{\scr \mu} u_{\scr L}^{\scr \beta}
  - \bar d^{}_{\scr L \alpha} \gamma^{\scr \mu} \nu_{\scr e L}^{\scr C}
  - \bar u^{}_{\scr R \alpha} \gamma^{\scr \mu} e_{\scr R}^{\scr +}
  \big) \\ \nonumber
& + & \dfrac{g^{}_{\scr U}}{\sqrt{2}} \; X^{\scr \alpha}_{\scr 3\mu} \; \big( 
   \bar d^{}_{\scr L \alpha} \gamma^{\scr \mu} e^{\scr -}_{\scr L}
  + \bar d^{}_{\scr R \alpha} \gamma^{\scr \mu} e^{\scr -}_{\scr R}
  + \bar u^{}_{\scr L \alpha} \gamma^{\scr \mu} \nu^{}_{\scr e L} 
  + \bar u^{}_{\scr R \alpha} \gamma^{\scr \mu} \nu^{}_{\scr e R} 
  \big) \\ \label{bnvld} 
& + & \textrm{h.c.}
\end{eqnarray}

Taking all possible fermion mixings into account one
obtains~\cite{ab} the effective four fermion Lagrangian 
\begin{eqnarray} \nonumber
\mathcal{L}_{\scr \textrm{ef\/f}}
& = & A^{}_1 \, 
\big( \varepsilon_{\scr \alpha \beta \gamma} \bar u_{\scr L}^{\scr C
    \gamma}  \gamma^{\scr \mu} u_{\scr L}^{\scr \beta} \big)
\big( \bar e^{\scr +}_{\scr L} \gamma_{\scr \mu} d_{\scr L}^{\scr \alpha} \big)
\; + \; A^{}_2 \, 
\big( \varepsilon_{\scr \alpha \beta \gamma} \bar u_{\scr L}^{\scr C
    \gamma}  \gamma^{\scr \mu} u_{\scr L}^{\scr \beta} \big)
\big( \bar e^{\scr +}_{\scr R} \gamma_{\scr \mu} d_{\scr R}^{\scr \alpha} \big)
\\ \nonumber
& + & A^{}_3 \, 
\big( \varepsilon_{\scr \alpha \beta \gamma} \bar u_{\scr L}^{\scr C
    \gamma}  \gamma^{\scr \mu} u_{\scr L}^{\scr \beta} \big)
\big( \bar \mu^{\scr +}_{\scr L} \gamma_{\scr \mu} d_{\scr L}^{\scr
  \alpha} \big) 
\; + \; A^{}_4 \, 
\big( \varepsilon_{\scr \alpha \beta \gamma} \bar u_{\scr L}^{\scr C
    \gamma}  \gamma^{\scr \mu} u_{\scr L}^{\scr \beta} \big)
\big( \bar \mu^{\scr +}_{\scr R} \gamma_{\scr \mu} d_{\scr R}^{\scr \alpha} 
\big) \\ \nonumber
& + & A^{}_5 \, 
\big( \varepsilon_{\scr \alpha \beta \gamma} \bar u_{\scr L}^{\scr C
    \gamma}  \gamma^{\scr \mu} u_{\scr L}^{\scr \beta} \big)
\big( \bar e^{\scr +}_{\scr L} \gamma_{\scr \mu} s_{\scr L}^{\scr \alpha} \big)
\; + \; A^{}_6 \, 
\big( \varepsilon_{\scr \alpha \beta \gamma} \bar u_{\scr L}^{\scr C
    \gamma}  \gamma^{\scr \mu} u_{\scr L}^{\scr \beta} \big)
\big( \bar e^{\scr +}_{\scr R} \gamma_{\scr \mu} s_{\scr R}^{\scr \alpha} 
\big) \\ \nonumber
& + & A^{}_7 \, 
\big( \varepsilon_{\scr \alpha \beta \gamma} \bar u_{\scr L}^{\scr C
    \gamma}  \gamma^{\scr \mu} u_{\scr L}^{\scr \beta} \big)
\big( \bar \mu^{\scr +}_{\scr L} \gamma_{\scr \mu} 
s_{\scr L}^{\scr \alpha} \big) 
\; + \; A^{}_8 \, 
\big( \varepsilon_{\scr \alpha \beta \gamma} \bar u_{\scr L}^{\scr C
    \gamma}  \gamma^{\scr \mu} u_{\scr L}^{\scr \beta} \big)
\big( \bar \mu^{\scr +}_{\scr R} \gamma_{\scr \mu} s_{\scr R}^{\scr \alpha} 
\big) \\ \nonumber
& + & A^{}_9 \, 
\big( \varepsilon_{\scr \alpha \beta \gamma} 
\bar u_{\scr L}^{\scr C \gamma} \gamma^{\scr \mu} d_{\scr L}^{\scr \beta} 
\big)
\big( \bar \nu_{\scr e R}^{\scr C} \gamma_{\scr \mu} d_{\scr R}^{\scr
  \alpha}
\big)
\; + \; A^{}_{10} \, 
\big( \varepsilon_{\scr \alpha \beta \gamma} 
\bar u_{\scr L}^{\scr C \gamma} \gamma^{\scr \mu} d_{\scr L}^{\scr \beta} 
\big)
\big( \bar \nu_{\scr \mu R}^{\scr C} \gamma_{\scr \mu} d_{\scr
  R}^{\scr \alpha} \big)
\\ \nonumber
& + & A^{}_{11} \, 
\big( \varepsilon_{\scr \alpha \beta \gamma} 
\bar u_{\scr L}^{\scr C \gamma} \gamma^{\scr \mu} d_{\scr L}^{\scr \beta} 
\big)
\big( \bar \nu_{\scr e R}^{\scr C} \gamma_{\scr \mu} s_{\scr
  R}^{\scr \alpha} \big)
\; + \; A^{}_{12} \, 
\big( \varepsilon_{\scr \alpha \beta \gamma} 
\bar u_{\scr L}^{\scr C \gamma} \gamma^{\scr \mu} d_{\scr L}^{\scr \beta} 
\big)
\big( \bar \nu_{\scr \mu R}^{\scr C} \gamma_{\scr \mu} s_{\scr R}^{\scr
  \alpha} \big)
\\ \nonumber
& + & A^{}_{13} \, 
\big( \varepsilon_{\scr \alpha \beta \gamma} 
\bar u_{\scr L}^{\scr C \gamma} \gamma^{\scr \mu} s_{\scr L}^{\scr \beta} 
\big)
\big( \bar \nu_{\scr e R}^{\scr C} \gamma_{\scr \mu} d_{\scr R}^{\scr
  \alpha} \big)
\; + \; A^{}_{14} \, 
\big( \varepsilon_{\scr \alpha \beta \gamma} 
\bar u_{\scr L}^{\scr C \gamma} \gamma^{\scr \mu} s_{\scr L}^{\scr \beta} 
\big)
\big( \bar \nu_{\scr \mu R}^{\scr C} \gamma_{\scr \mu} d_{\scr
  R}^{\scr \alpha} \big)
\\ \nonumber
& + & A^{}_{15} \, 
\big( \varepsilon_{\scr \alpha \beta \gamma} 
\bar u_{\scr L}^{\scr C \gamma} \gamma^{\scr \mu} d_{\scr L}^{\scr \beta} 
\big)
\big( \bar \nu_{\scr \tau R}^{\scr C} \gamma_{\scr \mu} d_{\scr R}^{\scr
  \alpha} \big)
\; + \; A^{}_{16} \, 
\big( \varepsilon_{\scr \alpha \beta \gamma} 
\bar u_{\scr L}^{\scr C \gamma} \gamma^{\scr \mu} d_{\scr L}^{\scr \beta} 
\big)
\big( \bar \nu_{\scr \tau R}^{\scr C} \gamma_{\scr \mu} s_{\scr
  R}^{\scr \alpha} \big)
\\ \nonumber
& + & A^{}_{17} \, 
\big( \varepsilon_{\scr \alpha \beta \gamma} 
\bar u_{\scr L}^{\scr C \gamma} \gamma^{\scr \mu} s_{\scr L}^{\scr \beta} 
\big)
\big( \bar \nu_{\scr \tau R}^{\scr C} \gamma_{\scr \mu} d_{\scr R}^{\scr
  \alpha} \big) \\ \nonumber
& + & \; \textrm{(\, terms with two $s$ quarks \,)}
\\ \nonumber
& + & \; \textrm{(\, terms with $c$ ,$b$ and $t$ quarks \,)}
\\ \nonumber
& + & \; \textrm{(\, terms with $\bar \tau^{\scr +}_{\scr L,R}$ and 
$\bar \nu^{\scr C}_{\scr e,\mu,\tau L}$ \,)}
\\ \label{eldfnd}
& + & \textrm{h.c.}
\end{eqnarray}
where the coefficients $A_{\scr i}$ are given as follows~\cite{cm}:
\begin{eqnarray*}
  A^{}_1 \; & = & \tilde G \; 
\big( ({\bf U}_{\scr R})_{11}^{} ({\bf U}_{\scr L})_{11}^{}
    + ({\bf U}_{\scr R})_{21}^{} ({\bf U}_{\scr L})_{21}^{} 
    + ({\bf U}_{\scr R})_{31}^{} ({\bf U}_{\scr L})_{31}^{} \big) \\
& & \hspace{0.4cm} \cdot
\big( ({\bf E}_{\scr R})_{11}^{} ({\bf D}_{\scr L})_{11}^{} 
    + ({\bf E}_{\scr R})_{21}^{} ({\bf D}_{\scr L})_{21}^{} 
    + ({\bf E}_{\scr R})_{31}^{} ({\bf D}_{\scr L})_{31}^{} \big)
\\
& + & \tilde G \; 
\big( ({\bf U}_{\scr R})_{11}^{} ({\bf D}_{\scr L})_{11}^{} 
+     ({\bf U}_{\scr R})_{21}^{} ({\bf D}_{\scr L})_{21}^{} 
+     ({\bf U}_{\scr R})_{31}^{} ({\bf D}_{\scr L})_{31}^{} \big) \\
& & \hspace{0.4cm} \cdot
\big( ({\bf E}_{\scr R})_{11}^{} ({\bf U}_{\scr L})_{11}^{} 
    + ({\bf E}_{\scr R})_{21}^{} ({\bf U}_{\scr L})_{21}^{} 
    + ({\bf E}_{\scr R})_{31}^{} ({\bf U}_{\scr L})_{31}^{} \big)
\\
A^{}_2 \; & = & \tilde G \; 
\big( ({\bf U}_{\scr R})_{11}^{} ({\bf U}_{\scr L})_{11}^{}
    + ({\bf U}_{\scr R})_{21}^{} ({\bf U}_{\scr L})_{21}^{} 
    + ({\bf U}_{\scr R})_{31}^{} ({\bf U}_{\scr L})_{31}^{} \big) \\
& & \hspace{0.4cm} \cdot
\big( ({\bf E}_{\scr L})_{11}^{} ({\bf D}_{\scr R})_{11}^{} 
    + ({\bf E}_{\scr L})_{21}^{} ({\bf D}_{\scr R})_{21}^{} 
    + ({\bf E}_{\scr L})_{31}^{} ({\bf D}_{\scr R})_{31}^{} \big)
\\
& + & \tilde G' \; 
\big( ({\bf D}_{\scr R})_{11}^{} ({\bf U}_{\scr L})_{11}^{} 
+     ({\bf D}_{\scr R})_{21}^{} ({\bf U}_{\scr L})_{21}^{} 
+     ({\bf D}_{\scr R})_{31}^{} ({\bf U}_{\scr L})_{31}^{} \big) \\
& & \hspace{0.5cm} \cdot
\big( ({\bf E}_{\scr L})_{11}^{} ({\bf U}_{\scr R})_{11}^{} 
    + ({\bf E}_{\scr L})_{21}^{} ({\bf U}_{\scr R})_{21}^{} 
    + ({\bf E}_{\scr L})_{31}^{} ({\bf U}_{\scr R})_{31}^{} \big)
\\
A^{}_3 \; & = & \tilde G \; 
\big( ({\bf U}_{\scr R})_{11}^{} ({\bf U}_{\scr L})_{11}^{}
    + ({\bf U}_{\scr R})_{21}^{} ({\bf U}_{\scr L})_{21}^{} 
    + ({\bf U}_{\scr R})_{31}^{} ({\bf U}_{\scr L})_{31}^{} \big) \\
& & \hspace{0.4cm} \cdot
\big( ({\bf E}_{\scr R})_{12}^{} ({\bf D}_{\scr L})_{11}^{}
    + ({\bf E}_{\scr R})_{22}^{} ({\bf D}_{\scr L})_{21}^{} 
    + ({\bf E}_{\scr R})_{32}^{} ({\bf D}_{\scr L})_{31}^{} \big)
\\
& + & \tilde G \; 
\big( ({\bf U}_{\scr R})_{11}^{} ({\bf D}_{\scr L})_{11}^{} 
+     ({\bf U}_{\scr R})_{21}^{} ({\bf D}_{\scr L})_{21}^{} 
+     ({\bf U}_{\scr R})_{31}^{} ({\bf D}_{\scr L})_{31}^{} \big) \\
& & \hspace{0.4cm} \cdot
\big( ({\bf E}_{\scr R})_{12}^{} ({\bf U}_{\scr L})_{11}^{} 
    + ({\bf E}_{\scr R})_{22}^{} ({\bf U}_{\scr L})_{21}^{} 
    + ({\bf E}_{\scr R})_{32}^{} ({\bf U}_{\scr L})_{31}^{} \big)
\\
A^{}_4 \; & = & \tilde G \; 
\big( ({\bf U}_{\scr R})_{11}^{} ({\bf U}_{\scr L})_{11}^{}
    + ({\bf U}_{\scr R})_{21}^{} ({\bf U}_{\scr L})_{21}^{} 
    + ({\bf U}_{\scr R})_{31}^{} ({\bf U}_{\scr L})_{31}^{} \big) \\
& & \hspace{0.4cm} \cdot
\big( ({\bf E}_{\scr L})_{12}^{} ({\bf D}_{\scr R})_{11}^{} 
    + ({\bf E}_{\scr L})_{22}^{} ({\bf D}_{\scr R})_{21}^{} 
    + ({\bf E}_{\scr L})_{32}^{} ({\bf D}_{\scr R})_{31}^{} \big)
\\
& + & \tilde G' \; 
\big( ({\bf D}_{\scr R})_{11}^{} ({\bf U}_{\scr L})_{11}^{} 
+     ({\bf D}_{\scr R})_{21}^{} ({\bf U}_{\scr L})_{21}^{} 
+     ({\bf D}_{\scr R})_{31}^{} ({\bf U}_{\scr L})_{31}^{} \big) \\
& & \hspace{0.5cm} \cdot
\big( ({\bf E}_{\scr L})_{12}^{} ({\bf U}_{\scr R})_{11}^{} 
    + ({\bf E}_{\scr L})_{22}^{} ({\bf U}_{\scr R})_{21}^{} 
    + ({\bf E}_{\scr L})_{32}^{} ({\bf U}_{\scr R})_{31}^{} \big)
\\
A^{}_5 \; & = & \tilde G \; 
\big( ({\bf U}_{\scr R})_{11}^{} ({\bf U}_{\scr L})_{11}^{}
    + ({\bf U}_{\scr R})_{21}^{} ({\bf U}_{\scr L})_{21}^{} 
    + ({\bf U}_{\scr R})_{31}^{} ({\bf U}_{\scr L})_{31}^{} \big) \\
& & \hspace{0.4cm} \cdot
\big( ({\bf E}_{\scr R})_{11}^{} ({\bf D}_{\scr L})_{12}^{} 
    + ({\bf E}_{\scr R})_{21}^{} ({\bf D}_{\scr L})_{22}^{} 
    + ({\bf E}_{\scr R})_{31}^{} ({\bf D}_{\scr L})_{32}^{} \big)
\\
& + & \tilde G \; 
\big( ({\bf U}_{\scr R})_{11}^{} ({\bf D}_{\scr L})_{12}^{} 
+     ({\bf U}_{\scr R})_{21}^{} ({\bf D}_{\scr L})_{22}^{} 
+     ({\bf U}_{\scr R})_{31}^{} ({\bf D}_{\scr L})_{32}^{} \big) \\
& & \hspace{0.4cm} \cdot
\big( ({\bf E}_{\scr R})_{11}^{} ({\bf U}_{\scr L})_{11}^{} 
    + ({\bf E}_{\scr R})_{21}^{} ({\bf U}_{\scr L})_{21}^{} 
    + ({\bf E}_{\scr R})_{31}^{} ({\bf U}_{\scr L})_{31}^{} \big)
\end{eqnarray*}
\pagebreak
\begin{eqnarray*}
A^{}_6 \; & = & \tilde G \; 
\big( ({\bf U}_{\scr R})_{11}^{} ({\bf U}_{\scr L})_{11}^{}
    + ({\bf U}_{\scr R})_{21}^{} ({\bf U}_{\scr L})_{21}^{} 
    + ({\bf U}_{\scr R})_{31}^{} ({\bf U}_{\scr L})_{31}^{} \big) \\
& & \hspace{0.4cm} \cdot
\big( ({\bf E}_{\scr L})_{11}^{} ({\bf D}_{\scr R})_{12}^{} 
    + ({\bf E}_{\scr L})_{21}^{} ({\bf D}_{\scr R})_{22}^{} 
    + ({\bf E}_{\scr L})_{31}^{} ({\bf D}_{\scr R})_{32}^{} \big)
\\
& + & \tilde G' \; 
\big( ({\bf D}_{\scr R})_{12}^{} ({\bf U}_{\scr L})_{11}^{} 
+     ({\bf D}_{\scr R})_{22}^{} ({\bf U}_{\scr L})_{21}^{} 
+     ({\bf D}_{\scr R})_{32}^{} ({\bf U}_{\scr L})_{31}^{} \big) \\
& & \hspace{0.5cm} \cdot
\big( ({\bf E}_{\scr L})_{11}^{} ({\bf U}_{\scr R})_{11}^{} 
    + ({\bf E}_{\scr L})_{21}^{} ({\bf U}_{\scr R})_{21}^{} 
    + ({\bf E}_{\scr L})_{31}^{} ({\bf U}_{\scr R})_{31}^{} \big)
\\
A^{}_7 \; & = & \tilde G \; 
\big( ({\bf U}_{\scr R})_{11}^{} ({\bf U}_{\scr L})_{11}^{}
    + ({\bf U}_{\scr R})_{21}^{} ({\bf U}_{\scr L})_{21}^{} 
    + ({\bf U}_{\scr R})_{31}^{} ({\bf U}_{\scr L})_{31}^{} \big) \\
& & \hspace{0.4cm} \cdot
\big( ({\bf E}_{\scr R})_{12}^{} ({\bf D}_{\scr L})_{12}^{} 
    + ({\bf E}_{\scr R})_{22}^{} ({\bf D}_{\scr L})_{22}^{} 
    + ({\bf E}_{\scr R})_{32}^{} ({\bf D}_{\scr L})_{32}^{} \big)
\\
& + & \tilde G \; 
\big( ({\bf U}_{\scr R})_{11}^{} ({\bf D}_{\scr L})_{12}^{} 
+     ({\bf U}_{\scr R})_{21}^{} ({\bf D}_{\scr L})_{22}^{} 
+     ({\bf U}_{\scr R})_{31}^{} ({\bf D}_{\scr L})_{32}^{} \big) \\
& & \hspace{0.4cm} \cdot
\big( ({\bf E}_{\scr R})_{12}^{} ({\bf U}_{\scr L})_{11}^{} 
    + ({\bf E}_{\scr R})_{22}^{} ({\bf U}_{\scr L})_{21}^{} 
    + ({\bf E}_{\scr R})_{32}^{} ({\bf U}_{\scr L})_{31}^{} \big)
\\
A^{}_8 \; & = & \tilde G \; 
\big( ({\bf U}_{\scr R})_{11}^{} ({\bf U}_{\scr L})_{11}^{}
    + ({\bf U}_{\scr R})_{21}^{} ({\bf U}_{\scr L})_{21}^{} 
    + ({\bf U}_{\scr R})_{31}^{} ({\bf U}_{\scr L})_{31}^{} \big) \\
& & \hspace{0.4cm} \cdot
\big( ({\bf E}_{\scr L})_{12}^{} ({\bf D}_{\scr R})_{12}^{} 
    + ({\bf E}_{\scr L})_{22}^{} ({\bf D}_{\scr R})_{22}^{} 
    + ({\bf E}_{\scr L})_{32}^{} ({\bf D}_{\scr R})_{32}^{} \big)
\\
& + & \tilde G' \; 
\big( ({\bf D}_{\scr R})_{12}^{} ({\bf U}_{\scr L})_{11}^{} 
+     ({\bf D}_{\scr R})_{22}^{} ({\bf U}_{\scr L})_{21}^{} 
+     ({\bf D}_{\scr R})_{32}^{} ({\bf U}_{\scr L})_{31}^{} \big) \\
& & \hspace{0.5cm} \cdot
\big( ({\bf E}_{\scr L})_{12}^{} ({\bf U}_{\scr R})_{11}^{} 
    + ({\bf E}_{\scr L})_{22}^{} ({\bf U}_{\scr R})_{21}^{} 
    + ({\bf E}_{\scr L})_{32}^{} ({\bf U}_{\scr R})_{31}^{} \big) 
\\
A^{}_9 \; & = & - \tilde G \; 
\big( ({\bf U}_{\scr R})_{11}^{} ({\bf D}_{\scr L})_{11}^{} 
+     ({\bf U}_{\scr R})_{21}^{} ({\bf D}_{\scr L})_{21}^{} 
+     ({\bf U}_{\scr R})_{31}^{} ({\bf D}_{\scr L})_{31}^{} \big) \\
& & \hspace{0.7cm} \cdot
\big( ({\bf N}_{\scr L})_{11}^{} ({\bf D}_{\scr R})_{11}^{} 
    + ({\bf N}_{\scr L})_{21}^{} ({\bf D}_{\scr R})_{21}^{} 
    + ({\bf N}_{\scr L})_{31}^{} ({\bf D}_{\scr R})_{31}^{} \big)
\\
& & - \tilde G' \; 
\big( ({\bf D}_{\scr R})_{11}^{} ({\bf D}_{\scr L})_{11}^{} 
+     ({\bf D}_{\scr R})_{21}^{} ({\bf D}_{\scr L})_{21}^{} 
+     ({\bf D}_{\scr R})_{31}^{} ({\bf D}_{\scr L})_{31}^{} \big) \\
& & \hspace{0.8cm} \cdot
\big( ({\bf N}_{\scr L})_{11}^{} ({\bf U}_{\scr R})_{11}^{} 
    + ({\bf N}_{\scr L})_{21}^{} ({\bf U}_{\scr R})_{21}^{} 
    + ({\bf N}_{\scr L})_{31}^{} ({\bf U}_{\scr R})_{31}^{} \big)
\\
A^{}_{10} & = & - \tilde G \; 
\big( ({\bf U}_{\scr R})_{11}^{} ({\bf D}_{\scr L})_{11}^{} 
+     ({\bf U}_{\scr R})_{21}^{} ({\bf D}_{\scr L})_{21}^{} 
+     ({\bf U}_{\scr R})_{31}^{} ({\bf D}_{\scr L})_{31}^{} \big) \\
& & \hspace{0.7cm} \cdot
\big( ({\bf N}_{\scr L})_{12}^{} ({\bf D}_{\scr R})_{11}^{} 
    + ({\bf N}_{\scr L})_{22}^{} ({\bf D}_{\scr R})_{21}^{} 
    + ({\bf N}_{\scr L})_{32}^{} ({\bf D}_{\scr R})_{31}^{} \big)
\\
& & - \tilde G' \; 
\big( ({\bf D}_{\scr R})_{11}^{} ({\bf D}_{\scr L})_{11}^{} 
+     ({\bf D}_{\scr R})_{21}^{} ({\bf D}_{\scr L})_{21}^{} 
+     ({\bf D}_{\scr R})_{31}^{} ({\bf D}_{\scr L})_{31}^{} \big) \\
& & \hspace{0.8cm} \cdot
\big( ({\bf N}_{\scr L})_{12}^{} ({\bf U}_{\scr R})_{11}^{} 
    + ({\bf N}_{\scr L})_{22}^{} ({\bf U}_{\scr R})_{21}^{} 
    + ({\bf N}_{\scr L})_{32}^{} ({\bf U}_{\scr R})_{31}^{} \big)
\\
A^{}_{11} & = & - \tilde G \; 
\big( ({\bf U}_{\scr R})_{11}^{} ({\bf D}_{\scr L})_{11}^{} 
+     ({\bf U}_{\scr R})_{21}^{} ({\bf D}_{\scr L})_{21}^{} 
+     ({\bf U}_{\scr R})_{31}^{} ({\bf D}_{\scr L})_{31}^{} \big) \\
& & \hspace{0.7cm} \cdot
\big( ({\bf N}_{\scr L})_{11}^{} ({\bf D}_{\scr R})_{12}^{} 
    + ({\bf N}_{\scr L})_{21}^{} ({\bf D}_{\scr R})_{22}^{} 
    + ({\bf N}_{\scr L})_{31}^{} ({\bf D}_{\scr R})_{32}^{} \big)
\\
& & - \tilde G' \; 
\big( ({\bf D}_{\scr R})_{12}^{} ({\bf D}_{\scr L})_{11}^{} 
+     ({\bf D}_{\scr R})_{22}^{} ({\bf D}_{\scr L})_{21}^{} 
+     ({\bf D}_{\scr R})_{32}^{} ({\bf D}_{\scr L})_{31}^{} \big) \\
& & \hspace{0.8cm} \cdot 
\big( ({\bf N}_{\scr L})_{11}^{} ({\bf U}_{\scr R})_{11}^{} 
    + ({\bf N}_{\scr L})_{21}^{} ({\bf U}_{\scr R})_{21}^{} 
    + ({\bf N}_{\scr L})_{31}^{} ({\bf U}_{\scr R})_{31}^{} \big)
\\
A^{}_{12} & = & - \tilde G \; 
\big( ({\bf U}_{\scr R})_{11}^{} ({\bf D}_{\scr L})_{11}^{} 
+     ({\bf U}_{\scr R})_{21}^{} ({\bf D}_{\scr L})_{21}^{} 
+     ({\bf U}_{\scr R})_{31}^{} ({\bf D}_{\scr L})_{31}^{} \big) \\
& & \hspace{0.7cm} \cdot
\big( ({\bf N}_{\scr L})_{12}^{} ({\bf D}_{\scr R})_{12}^{} 
    + ({\bf N}_{\scr L})_{22}^{} ({\bf D}_{\scr R})_{22}^{} 
    + ({\bf N}_{\scr L})_{32}^{} ({\bf D}_{\scr R})_{32}^{} \big)
\\
& & - \tilde G' \; 
\big( ({\bf D}_{\scr R})_{12}^{} ({\bf D}_{\scr L})_{11}^{} 
+     ({\bf D}_{\scr R})_{22}^{} ({\bf D}_{\scr L})_{21}^{} 
+     ({\bf D}_{\scr R})_{32}^{} ({\bf D}_{\scr L})_{31}^{} \big) \\
& & \hspace{0.8cm} \cdot 
\big( ({\bf N}_{\scr L})_{12}^{} ({\bf U}_{\scr R})_{11}^{} 
    + ({\bf N}_{\scr L})_{22}^{} ({\bf U}_{\scr R})_{21}^{} 
    + ({\bf N}_{\scr L})_{32}^{} ({\bf U}_{\scr R})_{31}^{} \big)
\\
A^{}_{13} & = & - \tilde G \; 
\big( ({\bf U}_{\scr R})_{11}^{} ({\bf D}_{\scr L})_{12}^{} 
+     ({\bf U}_{\scr R})_{21}^{} ({\bf D}_{\scr L})_{22}^{} 
+     ({\bf U}_{\scr R})_{31}^{} ({\bf D}_{\scr L})_{32}^{} \big) \\
& & \hspace{0.7cm} \cdot
\big( ({\bf N}_{\scr L})_{11}^{} ({\bf D}_{\scr R})_{11}^{} 
    + ({\bf N}_{\scr L})_{21}^{} ({\bf D}_{\scr R})_{21}^{} 
    + ({\bf N}_{\scr L})_{31}^{} ({\bf D}_{\scr R})_{31}^{} \big)
\\
& & - \tilde G' \; 
\big( ({\bf D}_{\scr R})_{11}^{} ({\bf D}_{\scr L})_{12}^{} 
+     ({\bf D}_{\scr R})_{21}^{} ({\bf D}_{\scr L})_{22}^{} 
+     ({\bf D}_{\scr R})_{31}^{} ({\bf D}_{\scr L})_{32}^{} \big) \\
& & \hspace{0.8cm} \cdot
\big( ({\bf N}_{\scr L})_{11}^{} ({\bf U}_{\scr R})_{11}^{} 
    + ({\bf N}_{\scr L})_{21}^{} ({\bf U}_{\scr R})_{21}^{} 
    + ({\bf N}_{\scr L})_{31}^{} ({\bf U}_{\scr R})_{31}^{} \big)
\\
A^{}_{14} & = & - \tilde G \; 
\big( ({\bf U}_{\scr R})_{11}^{} ({\bf D}_{\scr L})_{12}^{} 
+     ({\bf U}_{\scr R})_{21}^{} ({\bf D}_{\scr L})_{22}^{} 
+     ({\bf U}_{\scr R})_{31}^{} ({\bf D}_{\scr L})_{32}^{} \big) \\
& & \hspace{0.7cm} \cdot
\big( ({\bf N}_{\scr L})_{12}^{} ({\bf D}_{\scr R})_{11}^{} 
    + ({\bf N}_{\scr L})_{22}^{} ({\bf D}_{\scr R})_{21}^{} 
    + ({\bf N}_{\scr L})_{32}^{} ({\bf D}_{\scr R})_{31}^{} \big)
\\
& & - \tilde G' \; 
\big( ({\bf D}_{\scr R})_{11}^{} ({\bf D}_{\scr L})_{12}^{} 
+     ({\bf D}_{\scr R})_{21}^{} ({\bf D}_{\scr L})_{22}^{} 
+     ({\bf D}_{\scr R})_{31}^{} ({\bf D}_{\scr L})_{32}^{} \big) \\
& & \hspace{0.8cm} \cdot
\big( ({\bf N}_{\scr L})_{12}^{} ({\bf U}_{\scr R})_{11}^{} 
    + ({\bf N}_{\scr L})_{22}^{} ({\bf U}_{\scr R})_{21}^{} 
    + ({\bf N}_{\scr L})_{32}^{} ({\bf U}_{\scr R})_{31}^{} \big)
\end{eqnarray*}
\pagebreak
\begin{eqnarray*}
A^{}_{15} & = & - \tilde G \; 
\big( ({\bf U}_{\scr R})_{11}^{} ({\bf D}_{\scr L})_{11}^{} 
+     ({\bf U}_{\scr R})_{21}^{} ({\bf D}_{\scr L})_{21}^{} 
+     ({\bf U}_{\scr R})_{31}^{} ({\bf D}_{\scr L})_{31}^{} \big) \\
& & \hspace{0.7cm} \cdot
\big( ({\bf N}_{\scr L})_{13}^{} ({\bf D}_{\scr R})_{11}^{} 
    + ({\bf N}_{\scr L})_{23}^{} ({\bf D}_{\scr R})_{21}^{} 
    + ({\bf N}_{\scr L})_{33}^{} ({\bf D}_{\scr R})_{31}^{} \big)
\\
& & - \tilde G' \; 
\big( ({\bf D}_{\scr R})_{11}^{} ({\bf D}_{\scr L})_{11}^{} 
+     ({\bf D}_{\scr R})_{21}^{} ({\bf D}_{\scr L})_{21}^{} 
+     ({\bf D}_{\scr R})_{31}^{} ({\bf D}_{\scr L})_{31}^{} \big) \\
& & \hspace{0.8cm} \cdot
\big( ({\bf N}_{\scr L})_{13}^{} ({\bf U}_{\scr R})_{11}^{} 
    + ({\bf N}_{\scr L})_{23}^{} ({\bf U}_{\scr R})_{21}^{} 
    + ({\bf N}_{\scr L})_{33}^{} ({\bf U}_{\scr R})_{31}^{} \big)
\\
A^{}_{16} & = & - \tilde G \; 
\big( ({\bf U}_{\scr R})_{11}^{} ({\bf D}_{\scr L})_{11}^{} 
+     ({\bf U}_{\scr R})_{21}^{} ({\bf D}_{\scr L})_{21}^{} 
+     ({\bf U}_{\scr R})_{31}^{} ({\bf D}_{\scr L})_{31}^{} \big) \\
& & \hspace{0.7cm} \cdot
\big( ({\bf N}_{\scr L})_{13}^{} ({\bf D}_{\scr R})_{12}^{} 
    + ({\bf N}_{\scr L})_{23}^{} ({\bf D}_{\scr R})_{22}^{} 
    + ({\bf N}_{\scr L})_{33}^{} ({\bf D}_{\scr R})_{32}^{} \big)
\\
& & - \tilde G' \; 
\big( ({\bf D}_{\scr R})_{12}^{} ({\bf D}_{\scr L})_{11}^{} 
+     ({\bf D}_{\scr R})_{22}^{} ({\bf D}_{\scr L})_{21}^{} 
+     ({\bf D}_{\scr R})_{32}^{} ({\bf D}_{\scr L})_{31}^{} \big) \\
& & \hspace{0.8cm} \cdot 
\big( ({\bf N}_{\scr L})_{13}^{} ({\bf U}_{\scr R})_{11}^{} 
    + ({\bf N}_{\scr L})_{23}^{} ({\bf U}_{\scr R})_{21}^{} 
    + ({\bf N}_{\scr L})_{33}^{} ({\bf U}_{\scr R})_{31}^{} \big)
\\
A^{}_{17} & = & - \tilde G \; 
\big( ({\bf U}_{\scr R})_{11}^{} ({\bf D}_{\scr L})_{12}^{} 
+     ({\bf U}_{\scr R})_{21}^{} ({\bf D}_{\scr L})_{22}^{} 
+     ({\bf U}_{\scr R})_{31}^{} ({\bf D}_{\scr L})_{32}^{} \big) \\
& & \hspace{0.7cm} \cdot
\big( ({\bf N}_{\scr L})_{13}^{} ({\bf D}_{\scr R})_{11}^{} 
    + ({\bf N}_{\scr L})_{23}^{} ({\bf D}_{\scr R})_{21}^{} 
    + ({\bf N}_{\scr L})_{33}^{} ({\bf D}_{\scr R})_{31}^{} \big)
\\
& & - \tilde G' \;
\big( ({\bf D}_{\scr R})_{11}^{} ({\bf D}_{\scr L})_{12}^{} 
+     ({\bf D}_{\scr R})_{21}^{} ({\bf D}_{\scr L})_{22}^{} 
+     ({\bf D}_{\scr R})_{31}^{} ({\bf D}_{\scr L})_{32}^{} \big) \\
& & \hspace{0.8cm} \cdot
 \big( ({\bf N}_{\scr L})_{13}^{} ({\bf U}_{\scr R})_{11}^{} 
     + ({\bf N}_{\scr L})_{23}^{} ({\bf U}_{\scr R})_{21}^{} 
     + ({\bf N}_{\scr L})_{33}^{} ({\bf U}_{\scr R})_{31}^{} \big)
\end{eqnarray*}
We used here the definitions $\tilde G = g_{\scr{U}}^2/2M^2_{\scr X,Y}$ and
$\tilde G' = g_{\scr{U}}^2/2M^2_{\scr X',Y'}$, where
$M^2_{\scr X,Y} = M^2_{\scr X',Y'} \approx M^2_{\scr U}$ is assumed.

The coefficients $A_{\scr i}$ are connected to the hadronic transition 
amplitudes of the elementary processes responsible for the nucleon decays. 
The independent amplitudes are given in tables~\ref{aepd} and \ref{aend}.
\begin{table}[h]
\begin{center}
\begin{tabular}{|c|c|c|c|}
\hline
Decay process & Lagrangian term & Amplitude ($\cdot \sqrt{30}$)
& Coefficient \\
\hline \hline
$p \! \uparrow \; \rightarrow \;
e^{\scr +}_{\scr R} u^{\scr C} \!\! \uparrow u \! \downarrow$ 
& $\big( \bar u_{\scr L}^{\scr C}  \gamma^{\scr \mu} u_{\scr L}^{} \big)
\big( \bar e^{\scr +}_{\scr R} \gamma_{\scr \mu} d_{\scr R}^{} \big)$ 
& $-4$ & $A^{}_2$ \\
$p \! \uparrow \; \rightarrow \;
e^{\scr +}_{\scr R} u^{\scr C} \!\! \downarrow u \! \uparrow$ 
& $\big( \bar u_{\scr L}^{\scr C}  \gamma^{\scr \mu} u_{\scr L}^{} \big)
\big( \bar e^{\scr +}_{\scr R} \gamma_{\scr \mu} d_{\scr R}^{} \big)$ 
& $-8$ & $A^{}_2$ \\
$p \! \uparrow \; \rightarrow \;
e^{\scr +}_{\scr R} d^{\scr C} \!\! \uparrow d \! \downarrow$ 
& $\big( \bar u_{\scr L}^{\scr C}  \gamma^{\scr \mu} u_{\scr L}^{} \big)
\big( \bar e^{\scr +}_{\scr R} \gamma_{\scr \mu} d_{\scr R}^{} \big)$ 
& $-8$ & $A^{}_2$ \\
$p \! \uparrow \; \rightarrow \;
e^{\scr +}_{\scr R} d^{\scr C} \!\! \downarrow d \! \uparrow$ 
& $\big( \bar u_{\scr L}^{\scr C}  \gamma^{\scr \mu} u_{\scr L}^{} \big)
\big( \bar e^{\scr +}_{\scr R} \gamma_{\scr \mu} d_{\scr R}^{} \big)$ 
& $+2$ & $A^{}_2$ \\
\hline
$p \! \downarrow \; \rightarrow \;
e^{\scr +}_{\scr R} u^{\scr C} \!\! \downarrow u \! \downarrow$ 
& $\big( \bar u_{\scr L}^{\scr C}  \gamma^{\scr \mu} u_{\scr L}^{} \big)
\big( \bar e^{\scr +}_{\scr R} \gamma_{\scr \mu} d_{\scr R}^{} \big)$ 
& $-10$ & $A^{}_2$ \\
$p \! \downarrow \; \rightarrow \;
e^{\scr +}_{\scr R} d^{\scr C} \!\! \downarrow d \! \downarrow$ 
& $\big( \bar u_{\scr L}^{\scr C}  \gamma^{\scr \mu} u_{\scr L}^{} \big)
\big( \bar e^{\scr +}_{\scr R} \gamma_{\scr \mu} d_{\scr R}^{} \big)$ 
& $-2$ & $A^{}_2$ \\
\hline
$p \! \uparrow \; \rightarrow \;
e^{\scr +}_{\scr R} s^{\scr C} \!\! \uparrow d \! \downarrow$ 
& $\big( \bar u_{\scr L}^{\scr C}  \gamma^{\scr \mu} u_{\scr L}^{} \big)
\big( \bar e^{\scr +}_{\scr R} \gamma_{\scr \mu} s_{\scr R}^{} \big)$ 
& $-8$ & $A^{}_6$ \\
$p \! \uparrow \; \rightarrow \;
e^{\scr +}_{\scr R} s^{\scr C} \!\! \downarrow d \! \uparrow$ 
& $\big( \bar u_{\scr L}^{\scr C}  \gamma^{\scr \mu} u_{\scr L}^{} \big)
\big( \bar e^{\scr +}_{\scr R} \gamma_{\scr \mu} s_{\scr R}^{} \big)$ 
& $+2$ & $A^{}_6$ \\
\hline
$p \! \downarrow \; \rightarrow \;
e^{\scr +}_{\scr R} s^{\scr C} \!\! \downarrow d \! \downarrow$ 
& $\big( \bar u_{\scr L}^{\scr C}  \gamma^{\scr \mu} u_{\scr L}^{} \big)
\big( \bar e^{\scr +}_{\scr R} \gamma_{\scr \mu} s_{\scr R}^{} \big)$ 
& $-2$ & $A^{}_6$ \\
\hline
$p \! \uparrow \; \rightarrow \;
\nu_{\scr e R}^{\scr C} d^{\scr C} \!\! \uparrow u \! \downarrow$ 
& $\big( \bar u_{\scr L}^{\scr C}  \gamma^{\scr \mu} d_{\scr L}^{} \big)
\big( \bar \nu_{\scr e R}^{\scr C} \gamma_{\scr \mu} d_{\scr R}^{} \big)$ 
& $+4$ & $A^{}_9$ \\
$p \! \uparrow \; \rightarrow \;
\nu_{\scr e R}^{\scr C} d^{\scr C} \!\! \downarrow u \! \uparrow$ 
& $\big( \bar u_{\scr L}^{\scr C}  \gamma^{\scr \mu} d_{\scr L}^{} \big)
\big( \bar \nu_{\scr e R}^{\scr C} \gamma_{\scr \mu} d_{\scr R}^{} \big)$ 
& $-10$ & $A^{}_9$ \\
\hline
$p \! \downarrow \; \rightarrow \;
\nu_{\scr e R}^{\scr C} d^{\scr C} \!\! \downarrow u \! \downarrow$ 
& $\big( \bar u_{\scr L}^{\scr C}  \gamma^{\scr \mu} d_{\scr L}^{} \big)
\big( \bar \nu_{\scr e R}^{\scr C} \gamma_{\scr \mu} d_{\scr R}^{} \big)$ 
& $-8$ & $A^{}_9$ \\
\hline
$p \! \uparrow \; \rightarrow \;
\nu_{\scr e R}^{\scr C} s^{\scr C} \!\! \uparrow u \! \downarrow$ 
& $\big( \bar u_{\scr L}^{\scr C}  \gamma^{\scr \mu} d_{\scr L}^{} \big)
\big( \bar \nu_{\scr e R}^{\scr C} \gamma_{\scr \mu} s_{\scr R}^{} \big)$ 
& $+4$ & $A^{}_{11}$ \\
$p \! \uparrow \; \rightarrow \;
\nu_{\scr e R}^{\scr C} s^{\scr C} \!\! \downarrow u \! \uparrow$ 
& $\big( \bar u_{\scr L}^{\scr C}  \gamma^{\scr \mu} d_{\scr L}^{} \big)
\big( \bar \nu_{\scr e R}^{\scr C} \gamma_{\scr \mu} s_{\scr R}^{} \big)$ 
& $+2$ & $A^{}_{11}$ \\
$p \! \uparrow \; \rightarrow \;
\nu_{\scr e R}^{\scr C} s^{\scr C} \!\! \uparrow u \! \downarrow$ 
& $\big( \bar u_{\scr L}^{\scr C}  \gamma^{\scr \mu} s_{\scr L}^{} \big)
\big( \bar \nu_{\scr e R}^{\scr C} \gamma_{\scr \mu} d_{\scr R}^{} \big)$ 
& $0$ & $A^{}_{13}$ \\
$p \! \uparrow \; \rightarrow \;
\nu_{\scr e R}^{\scr C} s^{\scr C} \!\! \downarrow u \! \uparrow$ 
& $\big( \bar u_{\scr L}^{\scr C}  \gamma^{\scr \mu} s_{\scr L}^{} \big)
\big( \bar \nu_{\scr e R}^{\scr C} \gamma_{\scr \mu} d_{\scr R}^{} \big)$ 
& $-12$ & $A^{}_{13}$ \\
\hline
$p \! \downarrow \; \rightarrow \;
\nu_{\scr e R}^{\scr C} s^{\scr C} \!\! \downarrow u \! \downarrow$ 
& $\big( \bar u_{\scr L}^{\scr C}  \gamma^{\scr \mu} d_{\scr L}^{} \big)
\big( \bar \nu_{\scr e R}^{\scr C} \gamma_{\scr \mu} s_{\scr R}^{} \big)$ 
& $+4$ & $A^{}_{11}$ \\
$p \! \downarrow \; \rightarrow \;
\nu_{\scr e R}^{\scr C} s^{\scr C} \!\! \downarrow u \! \downarrow$ 
& $\big( \bar u_{\scr L}^{\scr C}  \gamma^{\scr \mu} s_{\scr L}^{} \big)
\big( \bar \nu_{\scr e R}^{\scr C} \gamma_{\scr \mu} d_{\scr R}^{} \big)$ 
& $-12$ & $A^{}_{13}$ \\
\hline
\end{tabular}
\end{center}
\caption{Decay amplitudes for the elementary processes of proton 
decays \label{aepd}}
\end{table}
\begin{table}
\begin{center}
\begin{tabular}{|c|c|c|c|}
\hline
Decay process & Lagrangian term & Amplitude ($\cdot \sqrt{30}$)
& Coefficient \\
\hline \hline
$n \! \uparrow \; \rightarrow \;
e^{\scr +}_{\scr R} u^{\scr C} \!\! \uparrow d \! \downarrow$ 
& $\big( \bar u_{\scr L}^{\scr C}  \gamma^{\scr \mu} u_{\scr L}^{} \big)
\big( \bar e^{\scr +}_{\scr R} \gamma_{\scr \mu} d_{\scr R}^{} \big)$ 
& $-4$ & $A^{}_2$ \\
$n \! \uparrow \; \rightarrow \;
e^{\scr +}_{\scr R} u^{\scr C} \!\! \downarrow d \! \uparrow$ 
& $\big( \bar u_{\scr L}^{\scr C}  \gamma^{\scr \mu} u_{\scr L}^{} \big)
\big( \bar e^{\scr +}_{\scr R} \gamma_{\scr \mu} d_{\scr R}^{} \big)$ 
& $+10$ & $A^{}_2$ \\
\hline
$n \! \downarrow \; \rightarrow \;
e^{\scr +}_{\scr R} u^{\scr C} \!\! \downarrow d \! \downarrow$ 
& $\big( \bar u_{\scr L}^{\scr C}  \gamma^{\scr \mu} u_{\scr L}^{} \big)
\big( \bar e^{\scr +}_{\scr R} \gamma_{\scr \mu} d_{\scr R}^{} \big)$ 
& $+8$ & $A^{}_2$ \\
\hline
$n \! \uparrow \; \rightarrow \;
\nu_{\scr e R}^{\scr C} u^{\scr C} \!\! \uparrow u \! \downarrow$ 
& $\big( \bar u_{\scr L}^{\scr C}  \gamma^{\scr \mu} d_{\scr L}^{} \big)
\big( \bar \nu_{\scr e R}^{\scr C} \gamma_{\scr \mu} d_{\scr R}^{} \big)$ 
& $+8$ & $A^{}_9$ \\
$n \! \uparrow \; \rightarrow \;
\nu_{\scr e R}^{\scr C} u^{\scr C} \!\! \downarrow u \! \uparrow$ 
& $\big( \bar u_{\scr L}^{\scr C}  \gamma^{\scr \mu} d_{\scr L}^{} \big)
\big( \bar \nu_{\scr e R}^{\scr C} \gamma_{\scr \mu} d_{\scr R}^{} \big)$ 
& $-2$ & $A^{}_9$ \\
$n \! \uparrow \; \rightarrow \;
\nu_{\scr e R}^{\scr C} d^{\scr C} \!\! \uparrow d \! \downarrow$ 
& $\big( \bar u_{\scr L}^{\scr C}  \gamma^{\scr \mu} d_{\scr L}^{} \big)
\big( \bar \nu_{\scr e R}^{\scr C} \gamma_{\scr \mu} d_{\scr R}^{} \big)$ 
& $+4$ & $A^{}_9$ \\
$n \! \uparrow \; \rightarrow \;
\nu_{\scr e R}^{\scr C} d^{\scr C} \!\! \downarrow d \! \uparrow$ 
& $\big( \bar u_{\scr L}^{\scr C}  \gamma^{\scr \mu} d_{\scr L}^{} \big)
\big( \bar \nu_{\scr e R}^{\scr C} \gamma_{\scr \mu} d_{\scr R}^{} \big)$ 
& $+8$ & $A^{}_9$ \\
\hline
$n \! \downarrow \; \rightarrow \;
\nu_{\scr e R}^{\scr C} u^{\scr C} \!\! \downarrow u \! \downarrow$ 
& $\big( \bar u_{\scr L}^{\scr C}  \gamma^{\scr \mu} d_{\scr L}^{} \big)
\big( \bar \nu_{\scr e R}^{\scr C} \gamma_{\scr \mu} d_{\scr R}^{} \big)$ 
& $+2$ & $A^{}_9$ \\
$n \! \downarrow \; \rightarrow \;
\nu_{\scr e R}^{\scr C} d^{\scr C} \!\! \downarrow d \! \downarrow$ 
& $\big( \bar u_{\scr L}^{\scr C}  \gamma^{\scr \mu} d_{\scr L}^{} \big)
\big( \bar \nu_{\scr e R}^{\scr C} \gamma_{\scr \mu} d_{\scr R}^{} \big)$ 
& $+10$ & $A^{}_9$ \\
\hline
$n \! \uparrow \; \rightarrow \;
\nu_{\scr e R}^{\scr C} s^{\scr C} \!\! \uparrow d \! \downarrow$ 
& $\big( \bar u_{\scr L}^{\scr C}  \gamma^{\scr \mu} d_{\scr L}^{} \big)
\big( \bar \nu_{\scr e R}^{\scr C} \gamma_{\scr \mu} s_{\scr R}^{} \big)$ 
& $+4$ & $A^{}_{11}$ \\
$n \! \uparrow \; \rightarrow \;
\nu_{\scr e R}^{\scr C} s^{\scr C} \!\! \downarrow d \! \uparrow$ 
& $\big( \bar u_{\scr L}^{\scr C}  \gamma^{\scr \mu} d_{\scr L}^{} \big)
\big( \bar \nu_{\scr e R}^{\scr C} \gamma_{\scr \mu} s_{\scr R}^{} \big)$ 
& $-4$ & $A^{}_{11}$ \\
$n \! \uparrow \; \rightarrow \;
\nu_{\scr e R}^{\scr C} s^{\scr C} \!\! \uparrow d \! \downarrow$ 
& $\big( \bar u_{\scr L}^{\scr C}  \gamma^{\scr \mu} s_{\scr L}^{} \big)
\big( \bar \nu_{\scr e R}^{\scr C} \gamma_{\scr \mu} d_{\scr R}^{} \big)$ 
& $0$ & $A^{}_{13}$ \\
$n \! \uparrow \; \rightarrow \;
\nu_{\scr e R}^{\scr C} s^{\scr C} \!\! \downarrow d \! \uparrow$ 
& $\big( \bar u_{\scr L}^{\scr C}  \gamma^{\scr \mu} s_{\scr L}^{} \big)
\big( \bar \nu_{\scr e R}^{\scr C} \gamma_{\scr \mu} d_{\scr R}^{} \big)$ 
& $+12$ & $A^{}_{13}$ \\
\hline
$n \! \downarrow \; \rightarrow \;
\nu_{\scr e R}^{\scr C} s^{\scr C} \!\! \downarrow d \! \downarrow$ 
& $\big( \bar u_{\scr L}^{\scr C}  \gamma^{\scr \mu} d_{\scr L}^{} \big)
\big( \bar \nu_{\scr e R}^{\scr C} \gamma_{\scr \mu} s_{\scr R}^{} \big)$ 
& $-2$ & $A^{}_{11}$ \\
$n \! \downarrow \; \rightarrow \;
\nu_{\scr e R}^{\scr C} s^{\scr C} \!\! \downarrow d \! \downarrow$ 
& $\big( \bar u_{\scr L}^{\scr C}  \gamma^{\scr \mu} s_{\scr L}^{} \big)
\big( \bar \nu_{\scr e R}^{\scr C} \gamma_{\scr \mu} d_{\scr R}^{} \big)$ 
& $+12$ & $A^{}_{13}$ \\
\hline
\end{tabular}
\end{center}
\caption{Decay amplitudes for the elementary processes of neutron
  decays \label{aend}}
\end{table}

\end{document}